%% file: main.tex
\documentclass[12pt,a4paper,english]{article} 
\usepackage[utf8]{inputenc}
\usepackage{babel}
\usepackage{sectsty}               
\usepackage{tabularx}              
\usepackage{longtable}
\usepackage{colortbl}              
\usepackage{pdflscape}
\usepackage{titling}               
\usepackage{imakeidx}              
\usepackage{xcolor}                
\usepackage{enumitem}              
\usepackage{tocloft}               
\usepackage{setspace}
\usepackage{graphicx}
\usepackage{color,soul}
\usepackage{tcolorbox}
\usepackage{threeparttable}
\usepackage{makecell}
\usepackage{amsmath}                                    
\usepackage{amssymb}                                    
\usepackage{listings}                                   
\usepackage{url}
\usepackage{comment}
\usepackage{caption}
\usepackage{algorithm}
\usepackage{algpseudocode}
\usepackage{subcaption}  
\DeclareCaptionFormat{algnote}{#1#2#3}
\newfloat{algonote}{htbp}{lon}
\floatname{algonote}{Note}

\usepackage[margin=1in]{geometry}
\usepackage{setspace}
\usepackage{booktabs}
\usepackage{adjustbox}

\usepackage{array}
\newcolumntype{H}{>{\setbox0=\hbox\bgroup}c<{\egroup}@{}}

\makeatletter
\newcommand\EightPtClose{\@setfontsize\EightPtClose\@viiipt{9}}
\newcommand\TenPtType{\@setfontsize\TenPtType\@xpt\@xiipt}
\def\notesize{\TenPtType}
\def\notesize{\EightPtClose}
\newenvironment{figurenotes}[1][\vspace{1em}Note]{\begin{minipage}[t]{\linewidth}\notesize{\itshape#1: }}{\end{minipage}}
\makeatother

\usepackage[colorinlistoftodos,prependcaption,textsize=tiny]{todonotes}
\usepackage{regexpatch}
\makeatletter
\xpatchcmd{\@todo}{\setkeys{todonotes}{#1}}{\setkeys{todonotes}{inline,#1}}{}{}
\makeatother

\let\oldmarginpar\marginpar
\renewcommand{\marginpar}[1]{\oldmarginpar{\tiny\color{red}#1}}

\usepackage[round]{natbib}
\definecolor{sangre}{RGB}{153, 0, 0}
\usepackage[xetex,hypertexnames=false,colorlinks,urlcolor = blue,citecolor =  sangre,linkcolor=blue]{hyperref}

\usepackage[normalem]{ulem}
\begin{document}

\title{The Benefits from Bundling Demand in K-12 Broadband Procurement\thanks{ We are grateful to Educational Services Commission of New Jersey for providing the data and to Kevin Dellicker, CEO of Dellicker Strategies, LLC  for helping us understand the market for K-12 broadband and the program. We also thank David Gibbons of Black Bear Fiber, Manudeep Bhuller, Yunmi Kong, Paulo Somaini, Rich Sweeney, Michael Whinston, Mo Xiao and conference and seminar participants at Boston College, Cornell, 2024 IIOC, HEC Montreal, the 2024 Conference on Auctions, Firm Behavior, and Policy, EC'24, UC Berkeley, the FCC, Helsinki GSE, JHU, UMD, 2025 Women in Empirical Microeconomics Conference, EARIE 2025 and 53rd TPRC 2025, for helpful comments and suggestions. The previous version of this paper was titled ``Bundling Demand in K-12 Broadband Procurement" and it appeared in the non-publication track at the 25th ACM Conference on Economics and Computation 2024 (EC'24).\\ {\bf Contact.} Aryal: \href{mailto:aryalg@bu.edu}{aryalg@bu.edu}, Murry: \href{mailto:ctmurry@umich.edu}{ctmurry@umich.edu}, Pal: \href{mailto:ppal2@stevens.edu}{ppal2@stevens.edu}, Palit: \href{mailto:arnab.palit@bateswhite.com}{arnab.palit@bateswhite.com}.}}

\author{
\begin{tabular}{@{}c@{\hspace{4em}}c@{}}
\textbf{Gaurab Aryal} & \textbf{Charles Murry} \\
{\small Boston University} & {\small University of Michigan \& NBER} \\[1em]
\textbf{Pallavi Pal} & \textbf{Arnab Palit} \\
{\small Stevens Institute of Technology} & {\small Bates White Economic Consulting}\\[-0.4em]
\small{School of Business}\\[1em]
\end{tabular}
}
\date{\today}  
\begingroup
\setlength{\droptitle}{-4em}
\maketitle
\endgroup

\vspace{-2em}  

\begin{abstract}

We study a new market design for K-12 school broadband procurement that switched from school-specific bidding to a system that bundled schools into groups. Using an event study approach, we estimate that the program reduced internet prices by \$9.17 (55\%) per Mbps per month while increasing bandwidth by 380.06 Mbps (136\%). These benefits resulted primarily from mitigating exposure risk in broadband procurement — the risk that providers win too few contracts to cover their fixed infrastructure costs. Using a bounds approach, we show robustness of our estimates and document that participants saved between \$1.61 million and \$3.48 million, while their existing federal E-rate subsidy was \$2.47 million, and experienced substantial welfare gains.

\end{abstract}

\newpage
\input{intro}

\input{model_and_mechanism}

\input{erate}

\input{data}

\input{analysis}

\input{welfare}

\input{conclusion}

\bibliographystyle{aer}
\bibliography{doe}

\appendix
\renewcommand{\thesection}{\Alph{section}} \setcounter{section}{0}
\renewcommand{\thesection}{\Alph{table}} \setcounter{table}{0}
\renewcommand{\thesubsection}{\Alph{section}.\arabic{subsection}} \setcounter{subsection}{0}
\renewcommand{\thesection}{\Alph{section}}
\renewcommand\thefigure{A.\arabic{figure}}    
\renewcommand\thetable{A.\arabic{table}}  
\input{appendix}

\end{document}

%% file: intro.tex

\section{Introduction}
In many auction settings, particularly in large-scale procurements, cost complementarities or returns to scale play an important role in determining outcomes because of the \emph{exposure problem} \citep{milgrom2004putting}. 
The exposure problem arises when there are complementarities or returns to scale across items, but each item is auctioned separately. 
In such cases, bidders cannot guarantee to win all complementary items, so they are ``exposed" to the risk of winning only a subset of items, leading to conservative bids, inefficiency, and lower revenue. 
A leading example of exposure is the FCC spectrum auction because carriers require adjacent frequencies and geographic regions for effective service.

Governments increasingly rely on large-scale procurements for private provision of public services (e.g., infrastructure, healthcare, and school lunches) that may be subject to exposure concerns.
Given that public procurement represents 12\%
of GDP in OECD countries, improving its efficiency could generate substantial economic benefits.
While auction theory has proposed \emph{package bidding} (or bundling) as a solution, its computational and strategic complexity \citep{RothkopfPekecHarstad1998}, significant administrative burdens \citep{KatokRoth2004}, and difficulty in maintaining transparency \citep{AusubelMilgrom2002} limit practical use.\footnote{There is a long literature examining the trade-off between optimal and simple mechanisms. For recent advances in characterizing simple, practical mechanisms, see, e.g., \cite{Li2017} and \cite{PyciaTroyan2023}.}
Moreover, there is limited direct empirical evidence on the benefits of design changes, such as bundling, on procurement outcomes. 

In this paper, we provide direct evidence using novel data on the outcomes following a design change to K-12 broadband internet procurement, a setting with clear exposure concerns.
In the U.S., the federal government subsidizes K-12 schools' internet contracts through a program called \emph{E-Rate}.
To receive the subsidy, schools must use competitive bidding to choose internet service providers (ISPs), and each school typically organizes its auction individually. 
In 2014, the New Jersey Department of Education initiated a program that bundled willing schools into four regional groups, requiring providers to bid for entire regions collectively (pure bundles). This design is less complex than package bidding solutions used in other contexts \citep[e.g.,][]{FoxBajari2013, KimOlivaresWeintraub2014}.
We evaluate the price and bandwidth benefits of this design change.

Our empirical strategy compares contracts from 2014, before the program began, to those from 2015, the first year after the program began. 
In particular, we use a difference-in-differences (DiD) approach and estimate the average treatment effect of bundling on internet prices and bandwidth for the participating schools.
Under the standard DiD assumptions, we estimate the causal effect of the program to be a \$9.37 per megabit per second (Mbps) per month price decrease (from a base of \$16.58) and a 308.54 Mbps increase (from a base of 278.09 Mbps) in the internet speed or bandwidth. 
Thus, a streamlined pure bundling that offers fewer but strategically chosen package combinations improves prices and speed, providing new empirical evidence of the benefits of practical market design solutions in public procurement.

Two mechanisms rationalize the sharp price decline: mitigation of the exposure problem and increased competition for larger (bundled) contracts.
First, ISPs face large fixed costs to build broadband infrastructure, which is built out in a network structure with hubs and spokes. 
Telecommunication markets have the feature that contiguous service areas are cheaper for providers to serve than dispersed ones \citep{AusubelCramtonMcAfeeMcMillan1997, Beresteanu2005,elliott2023market}.  
Under decentralized procurement, ISPs risk winning contracts for only one school in a geographic area, which may lead them to place higher bids than they would if they could bid on a bundle of schools in the same area.
Second, by grouping schools, the program may have induced additional competition for the bundle than for individual schools, which can also lead to lower prices.

To formalize the intuition behind the first mechanism (exposure), we present a stylized model with a simulation exercise demonstrating that pure bundling \emph{can} reduce procurement prices. We also separately analyze the effect of bundling on two distinct internet services offered by the Consortium, which differ substantially in their fixed costs and infrastructure requirements.  
The first is a basic single-connection service that utilizes existing infrastructure, typically suitable for individual schools or small districts. The second, the most commonly demanded service, is a complex dedicated enterprise solution requiring significant infrastructure investment and regional hubs.
Given these characteristics, the basic service faces minimal exposure risk, while the enterprise solution is particularly vulnerable to exposure. Our empirical results are consistent with this distinction: We find no significant price effects for the basic service but observe a substantial decrease of approximately \$9.17 per Mbps for the enterprise solution. 

For the second mechanism (competition), we control for competition in our DiD specification using the number of ISPs present in each school's area. The coefficient estimate on this measure is too small to explain the estimated price reductions. We further explore this issue using alternative competition measures based on the number of bids submitted in the Consortium, considering both conservative and generous assumptions about competition. Even under the most favorable scenario for competition effects, we estimate that the Consortium leads to a substantial decrease in price, confirming that exposure effects dominate. 
While the level of competition may be endogenous, our findings indicate that mitigating exposure risk through bundling, rather than increased competition, is the primary mechanism that generates the observed benefits from bundling.

We also evaluate the effect of participating on schools' expenditures and welfare. 
Our estimates suggest substantial yearly cost savings to participating schools, ranging from \$1.61 million to \$3.48 million (in total) for participants who purchased the enterprise solution, depending on assumptions about their bandwidth choices in the absence of the program.  
In comparison, participating schools that purchased the enterprise solution received a total of \$2.47 million in E-rate subsidies the year before joining the program. 

Next, we determine how these savings translate into changes in schools' welfare. 
Typically, quantifying welfare change requires estimating schools' demand. 
Instead, we follow the ``robust bounds approach'' of \cite{kang2022robust} and determine the lower and upper bounds for the \emph{change in welfare} from 2014 to 2015.
Under the assumption that the demand functions are log-concave, the change in welfare due to bundling is positive and large, suggesting widespread benefits for schools.

Therefore, bundling demand delivers substantial savings to schools and the government. 
The federal government has identified fast and affordable Internet as a policy goal \citep{Infrastructure2021}, and practically, the FCC has been concerned about the ballooning costs of supporting schools’ broadband needs. 
One of the three goals of the 2014 E-rate Modernization Order \citep{FCC2014} is 
promoting practices enabling schools and libraries to get the most out of the subsidies provided.
Our results are important because they identify demand bundling as a potentially effective approach to achieving this dual goal.

Participation in the program is voluntary, and only some New Jersey schools participate. 
We may be concerned about the assumptions required to estimate treatment effects with the DiD approach.
To address this concern, we explore how sensitive our estimates are to the assumption of parallel trends.
Specifically, we use the insights of \cite{manski2018right} and determine the treatment effects as a function of the degree of violation in the parallel trends assumption.
Our results are robust to violations in parallel trends. 
In particular, we find that the treatment group trend must be about 2.5 times or more than 2.5 times steeper than the control group trend to erase our treatment effect on the price and bandwidth. 
Furthermore, as we explain later, there are economic reasons to expect the trend to be steeper for the control group than for the treatment group, suggesting that our findings are robust to selection concerns. 

We also perform additional robustness checks. First, some schools may not be appropriate as controls if they were locked into multi-year contracts in 2014. 
Second, some nonparticipating schools may be located near participating schools and thus have access to lower prices from winning providers.
We find similar results after removing both types of schools from the control group. 
Additionally, to understand the program's potential total impact, we also examine effects on control schools by computing the difference in price between their current contract and the winning contract in their geographic region. Sixty-nine schools signed new contracts in 2015, but with non-winning ISPs. If we use only these schools as the control group, we find that bundling decreases price by \$6.69 per Mbps.  
Finally, while bundling could violate SUTVA through infrastructure spillovers that affect both participants and non-participants, these effects likely take several years to manifest and are not reflected in our data.

\subsubsection*{Related Literature}

Our main contributions are to the empirical auction design literature that studies the benefits and proper design of auctions.
Many papers \citep[e.g.,][]{BajariMcMillanTadelis2009,roberts2013should, CovertSweeney2023, DingDugganStarc2025} have confirmed theoretical insights by empirically documenting the benefits of auctions. 
Our paper is the first to document the effect of pure bundling using field data. 
This result is important because others, using structural empirical methods, have shown that complementarities may play an important role in auction outcomes and that bundling or package bidding can significantly improve the efficiency of auctions, for example, see \citet{CantillonPesendorfer2010, CapliceSheffi2010} for auctions of transportation services, \citet{FoxBajari2013, xiao2022license} for demand complementarities in spectrum auctions, \citet{KimOlivaresWeintraub2014, AgarwalLiSomaini2023} for procurements of school lunches, and \cite{GentryKomarovaSchiraldi2023} for the economy of scale in procurements of highway construction services.\footnote{There is related work that documents the exposure problem and the benefits of package bidding using laboratory experiments  \citep[e.g.,][]{KagelLienMilgrom2010, GoereeLindsay2019}.}

Our results will likely inform policymakers in other private-public settings, where essential services are provided by the private sector but paid for by public funds. Examples include healthcare, social security, and transportation.
In recent years, policymakers across various sectors have employed market design tools to improve the outcomes of such public-private partnerships \citep{CBO2017, CBO2020}.
As \cite{DecarolisPolyakovaRyan2020} points out, the prima facie objective of using private markets in combination with public provision of such services is to ``leverage the benefits of competition to provide high-quality services at low cost to both consumers and the government.''
Our results highlight that when such partnerships are structured to leverage demand aggregation and bundling, they can effectively enhance service delivery while minimizing costs. 
These findings underscore the broader applicability of our results, suggesting that carefully crafted procurement strategies can optimize outcomes not just in isolated cases but across a range of essential services where public funds are involved.

Lastly, we contribute to the policy literature on Internet access by understanding ways to use insights from market design to improve access and affordability of technological resources for K-12 students. 
\cite{goolsbee2006impact} show that the E-rate program significantly improved California schools' internet access in the late 1990s. Recent federal initiatives continue to prioritize school connectivity, including the Emergency Connectivity Fund and Digital Equity Act Programs in the 2021 Infrastructure Investment and Jobs Act. However, their effectiveness remains to be evaluated. Our contribution is to document that bundling can complement these programs and improve internet access.

%% file: model_and_mechanism.tex

\section{Model}
\label{sec:model}

In this section, we present a stylized model of procurement with economies of scale. 
The goal is to capture the exposure mechanism through which bundling can lead to lower payments. We begin by outlining the basic setup, then describe decentralized procurement, followed by centralized procurement with pure bundling. The main result is that the total payments under pure bundling \emph{can be} lower than the total payments under decentralized bidding when the economies of scale are sufficiently large.  

\subsubsection*{Environment}
Suppose there are two schools, indexed by $s \in S= \{1,2\}$. Initially, these schools conduct auctions for the Internet separately. Suppose there are $N\geq2$ ISPs who are interested in servicing both schools. Let $c_{is}\in [\underline{c}, \overline{c}]$ denote ISP $i$'s cost of providing internet service to school $s$. We assume that these costs are independently and identically distributed (across ISPs and schools) as $F_c(\cdot)$ with density $f_c(\cdot)>0$. 

Furthermore, suppose that if ISP $i$ services both schools, then its total cost is 
\begin{equation}
    \varphi(c_{i1}, c_{i2})= (c_{i1}+c_{i2}) - \Gamma,
\end{equation} 
where $0\leq \Gamma\leq 2\underline{c}$ captures cost savings from serving both schools together.   While $\Gamma$ could depend on the observed characteristics of the schools, including the terrain, and may also be bidder-specific, for parsimony, we assume that it is deterministic and common for all bidders. When it is clear from the context, for notational ease, we use $\varphi_i$ to denote the total cost $\varphi(c_{i1}, c_{i2})$, and let $F^*(\cdot)$ be its distribution, which is given by an appropriate convolution of $F_c(\cdot)$, with support $[\underline{\varphi}=\varphi(\underline{c}, \underline{c}), \overline{\varphi}=\varphi(\overline{c}, \overline{c})]$.  

\subsubsection*{Decentralized Procurement} 
Each school chooses the winning ISP as the one that submits the lowest bid among all the bids received by the school. For simplicity, we assume that all $N$ ISPs submit bids for both schools simultaneously. 
ISP $i$, with costs $(c_{i1}, c_{i2})$, chooses bids $(b_{i1}, b_{i2})$ to maximize its expected profit:
\begin{align}
    \max_{(b_1, b_2)}\Big\{ \sum_{s=1}^2(b_s - c_{is}) &\times \Pr(\text{win school $s$ and lose $s'\neq s$}) \notag \\
    & + (b_1 + b_2 - \varphi_i) \times \Pr(\text{win both schools})\Big\}.
    \end{align}
Bidders cannot condition their bid for a school on the outcome of the other school, and because schools decide the winner independently, $\Pr(\text{win both schools})=\prod_s\Pr(\text{win school s})$. This feature of decentralized bidding is the source of the exposure effect. 

We can rewrite the objective function to clarify the benefit of serving both schools as: 
\begin{align}
\max_{(b_1, b_2)}  \sum_{s=1}^2\Bigg((b_s - c_{is}) +\frac{\Gamma}{2} \times \Pr(\text{win school $s'\neq s$})\Bigg)\times \Pr(\text{win school $s$}).
\end{align}
We focus on a symmetric (and exchangeable) bidding strategy $\beta: [\underline{c}, \overline{c}]\times [\underline{c}, \overline{c}] \rightarrow \mathbb{R}_+$.
Under the assumption that $(N-1)$ bidders use $\beta$ to choose their bids, an ISP $i$ with costs $(c_{is}, c_{is'})$ chooses bids $(b_s, b_s')$ that maximizes its expected profit, i.e., 
\begin{align}
\max_{(b_s, b_s')} \sum_{s=1}^2\Big((b_s - c_{is}) +\frac{\Gamma}{2} \times (1-F_c(\beta^{-1}(b_{s'})))^{(N-1)}\Big)\times (1-F_c(\beta^{-1}(b_s)))^{(N-1)},
\end{align}
where with a slight abuse of notation, we use $\beta(c_{is})$ to denote the bid submitted by an ISP for school $s$ with cost $c_{is}$ after suppressing the dependence of the bids on $i$'s cost of serving school $s', c_{is'}$. 
The optimal bidding function in the decentralized system, which we denote by $\beta^{\text{pre}}$, can be written as follows:
\begin{align}
\beta^{\text{pre}}(c_{is}, c_{is'})=\underbrace{c_{is}}_{\text{stand alone cost}} + \underbrace{\int_{c_{is}}^{\overline{c}} \left(\frac{1-F_c(t)}{1-F_c(c_{is})}\right)^{(N-1)} dt}_{\text{stand alone markup}}-\underbrace{\Gamma \times(1-F_c(c_{is'}))^{(N-1)}}_{\text{expected cost savings}}.\label{eq:bid_pre}
\end{align}
Thus, even under decentralized bidding, the expected cost savings of serving both schools are reflected in bids, as ISPs pass these savings to remain competitive.  
The bidding strategy nests, as a special case, the bidding strategy in a standard procurement when $\Gamma=0$. Adding (\ref{eq:bid_pre}) for both schools gives the total bid under decentralized procurement. 

\subsubsection*{Pure Bundling} 
We now consider a centralized procurement model in which the two schools are bundled, and ISPs compete for both.
They cannot bid for only one school. Thus, this system can be best understood as ``pure bundling"  because ISPs cannot bid for only one school. 
Under pure bundling, ISP $i$ with costs $(c_{i1}, c_{i2})$ (and total cost $\varphi_i=\varphi(c_{i1}, c_{i2})$) chooses a total bid $b$ to maximize its expected profit:
\begin{align}
\max_{b} \Big\{(b - \varphi_i) \times  \Pr(\text{win})\Big\}.
\end{align}
This maximization problem is a standard procurement problem for a single item where the cost is $\varphi_i\sim F^*(\cdot)$.  
Therefore, the equilibrium bidding strategy under bundling is 
\begin{align}
\beta^{\text{post}}(\varphi_i) = \underbrace{\varphi_i}_{\text{total cost}} + \underbrace{\int_{\varphi_i}^{\overline{\varphi}} \left(\frac{1-F^*(t)}{1-F^*(\varphi_i)}\right)^{(N-1)} dt}_{\text{total markup}}.\quad\label{eq:bid_post}
\end{align}

\subsubsection*{Analysis: The Benefits from Bundling} 
The theoretical literature on bundling in procurements shows that the outcomes depend critically on the interplay between cost complementarities, the number of bidders, and the distribution of costs \citep{KrishnaRosenthal1996, Chakraborty1999, AveryHendershott2000, SubramaniamVenkatesh2009}. Establishing general conditions under which bundling \emph{always} reduces procurement costs requires characterizing the complex relationship between the order statistics of individual costs and their sum, which varies with the complementarity parameter ($\Gamma$), the number of bidders, and the number of schools (or the size of the bundle). 
Given that such a characterization is a hard problem, we use simulations to illustrate the key mechanism that is the most relevant for our context: bundling allows firms to internalize cost complementarities directly in their bidding strategies, reducing exposure to the risk of winning only one contract when efficiencies arise from serving multiple contracts simultaneously.

Our simulation design captures some key features of broadband procurement settings. We consider procurements with varying numbers ($N\in\{2,3,5,10\}$) of bidding internet service providers (ISPs) competing to provide broadband services to two schools, $s \in S = \{1,2\}$. We assume that costs for serving individual schools follow truncated Exponential distributions with mean $40$ and support $[10, 100]$, with costs independent across bidders and schools. We also vary the complementarity parameter ($\Gamma\in\{4, 8, 12, 16\}$), which represents the cost savings an ISP realizes when serving both schools simultaneously rather than individually. These parameter values generate cost savings ranging from 10\% to 40\% of the expected individual school costs, reflecting realistic economies of scope in infrastructure deployment. For each combination of $N$ and $\Gamma$, we simulate $1,000$ auctions under both decentralized and bundled procurement regimes.

\begin{figure}[t!!]
\caption{Bids with and without Bundling}\label{fig:bids_comparision}
\begin{subfigure}[b]{0.4\textwidth}
    \includegraphics[width=\textwidth]{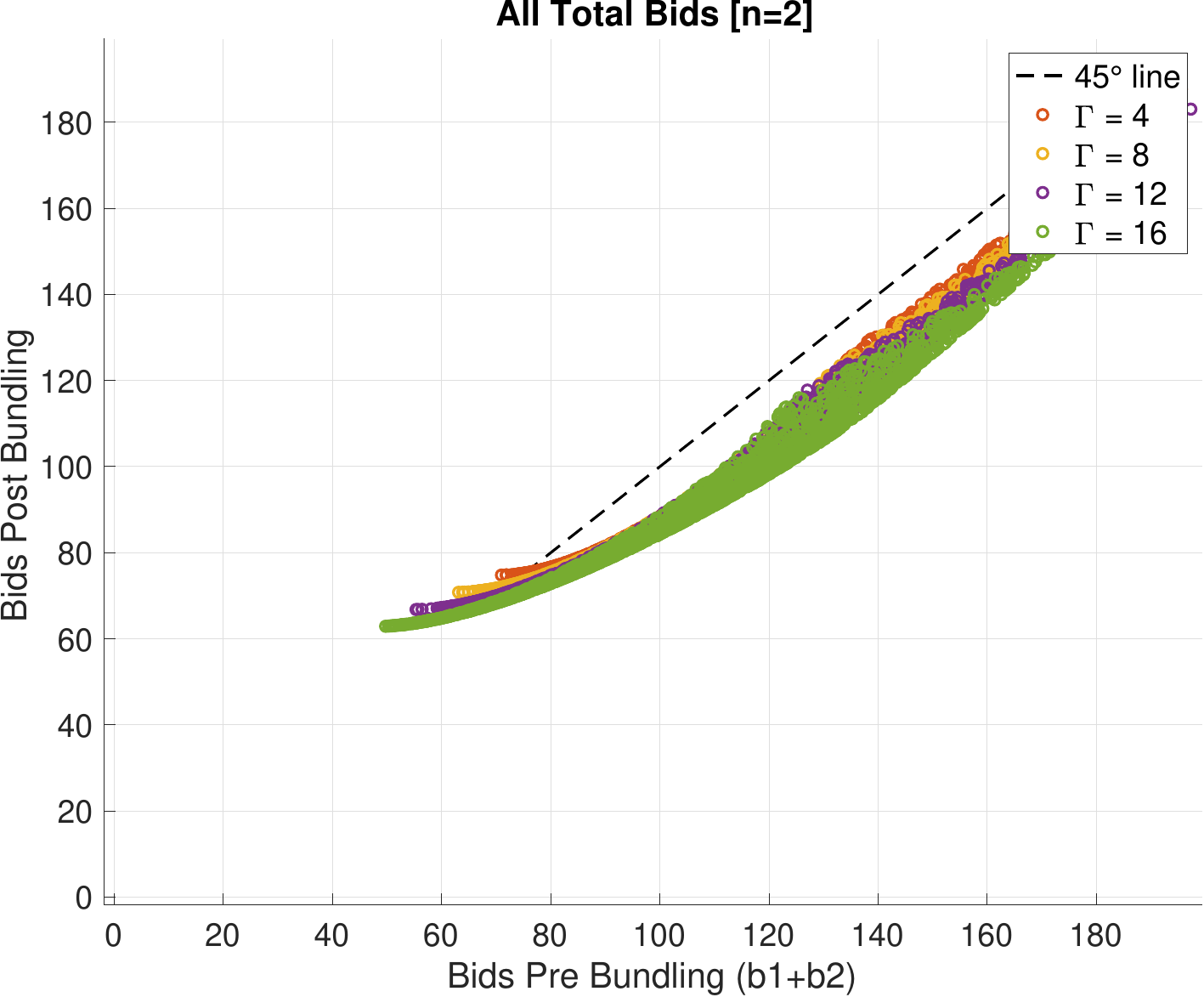}
    \caption{ All Bids}
    \label{fig:all_bids}
\end{subfigure}
\hfill
\begin{subfigure}[b]{0.4\textwidth}
    \includegraphics[width=\textwidth]{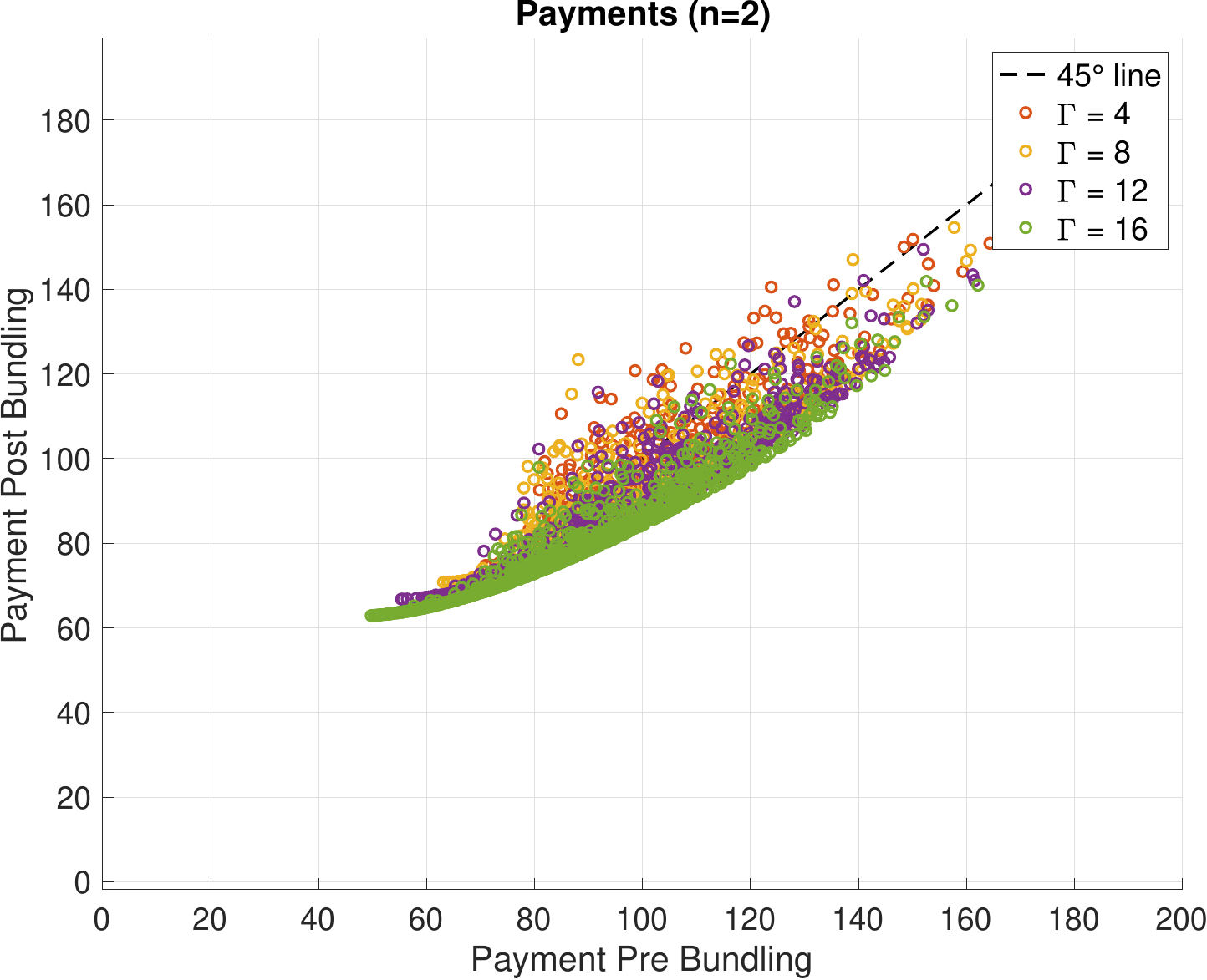}
    \caption{Final Payments}
    \label{fig:winning_bids}
\end{subfigure}
\begin{figurenotes}
The figure compares the bids before and after the bundling. Subfigure (a) shows the scatterplot of total bids before bundling (x-axis) and the total bids post bundling (y-axis) for fixed $N=2$ bidders and different values of $\Gamma$, and (b) shows the corresponding total payments before bundling (x-axis) and post bundling (y-axis). 
\end{figurenotes}
\end{figure}

Figure \ref{fig:bids_comparision} presents our simulation results for the case of $N=2$ ISPs, which provides the clearest illustration of the bundling effect. Panel (a) displays total bids for each ISP under decentralized bidding (by summing Equation (\ref{eq:bid_pre}) across schools) in the x-axis, and the bundle bids in the y-axis, and Panel (b) is the scatter plot of winning bids. The predominant pattern shows scatter points below the 45-degree line, confirming that bundling reduces total bids and total payments, by an amount that increases with $\Gamma$.\footnote{While we find that bundling generally reduces total procurement costs for our parameterization, occasionally total pre-bundling bids can be lower than post-bundling (see the southwest region in Figure \ref{fig:bids_comparision}(a)). This low-probability phenomenon occurs when ISPs draw low costs for \emph{both} schools, creating a ``double-counting'' effect: bidders reduce each school's bid by nearly the full complementarity parameter $\Gamma$ (anticipating winning the other school, as shown in Equation (\ref{eq:bid_pre})), resulting in lower total bids.}

Furthermore, all else being equal, as the number of bidders increases, the competition leads to a more pronounced decrease in bids. 
Table \ref{tab:average_bids} presents the average total bids and payments for different $N$ and $\Gamma$  parameters. The averages are taken over the simulations and the bidders for each simulation. Notably, the average bids decrease with $N$ and $\Gamma$. 

The economic intuition underlying these results is straightforward: under decentralized procurement, ISPs face exposure risk—they may win one school contract but lose the other, foregoing potential cost complementarities. This risk is reflected in more conservative bidding strategies. Bundling eliminates this exposure risk by ensuring that winners serve both schools and can fully realize cost efficiencies, leading to more aggressive bidding and lower procurement costs. 
\begin{table}[ht!!]
\centering
\caption{Effect of Bundling on Average Bids and Payments}\label{tab:average_bids}
	\begin{threeparttable}
	\begin{tabular}{ccccccccc}
\toprule
&  \multicolumn{4}{c}{ All Total Bids} & \multicolumn{4}{c}{Final Payments } \\
$N\backslash\Gamma$ & 4 & 8 & 12 & 16 & 4 & 8 & 12 & 16  \\
\hline
2 &  112.14&	98.07&	87.65&	82.22&  96.55&	72.47&	51.94&	38.13 \\
3 &  108.28&	94.75&	86.08&	79.18&  92.69&	68.54&	49.89&	35.35 \\
5 &  104.54&	91.04&	81.80&	77.28&  89.26&	64.50&	46.24&	32.34\\
10  & 99.64&	88.41&	80.18&	75.51&  83.97&	62.44&	43.49&	29.84 \\
\bottomrule
\end{tabular}
		\begin{tablenotes}[flushleft]
		\item \footnotesize \emph{Note:} The table presents the average total bids and the average final payments for $(N,\Gamma)\in \{2, 3, 5, 10 \}\times \{4, 8, 12, 16\}$. These bids and payments are averaged across bidders and $1,000$ simulations. 
		\end{tablenotes}
		\end{threeparttable}
\end{table}

%% file: erate.tex
\section{Institutional Details}
\label{sec:setting}

Our setting involves the procurement of high-speed broadband internet by public and private school districts and libraries (henceforth, schools) in the U.S. state of New Jersey.\footnote{``Broadband'' is a generic term for high-capacity transmissions, such as fiber optic or coaxial wires.}
New Jersey's Digital Readiness for Learning and Assessment Project (DRLAP) was launched in 2013 by the state's Department of Education to help K-12 schools better incorporate technology into their classrooms. The broadband component of the program, known as DRLAP-Broadband, was designed to help schools work together to improve their internet access and network infrastructure to bridge the technology gap across schools and ensure internet access necessary to utilize new digital resources.
	
Typically and exclusively before 2014, schools in New Jersey organized the procurement of internet and other telecommunications services individually. In 2014, New Jersey began centralizing the procurement process to reduce costs and increase access to high-speed internet. The design change was meant to meet the need for federal internet subsidies. Below, we describe how subsidies work and the particular intervention in New Jersey.

\subsection{The E-rate Program}
In the U.S., K-12 schools can apply for subsidies for their internet expenses through a federally funded program called {E-rate}, which is administered by the FCC and funded by the \emph{Universal Service Fund} under the \emph{Telecommunications Act of 1996}. The subsidy ranges from 20\% to 90\% of a school's telecommunications expenditures, depending on the poverty level of its students and its rural status. The E-rate program was designed to help eligible schools obtain internet.	In particular, the FCC set a goal of 1 Mbps per student to support digital learning in every classroom. The total subsidy cap in 2023 was \$4.5 billion. In 2023, 74\% of school districts met this goal, compared to 8\% in 2015. 

The typical procurement and subsidy process is decentralized.
A school determines the amount of internet it needs (e.g., unlimited internet with 1,000 Mbps download speed) and submits a request for competitive bids to the Universal Service Administrative Company (USAC) by filing FCC Form 470. USAC posts these requests on its website, and interested ISPs submit bids. After reviewing the bids, the school selects the most cost-effective ISP and files FCC Form 471 with details of the chosen ISP, following which either the school or the chosen ISP can apply to USAC for reimbursement. All eligible schools that conduct a fair and open competitive bidding process receive subsidies.\footnote{See \url{https://e-ratecentral.com/Resources/Educational-Information/The-E-Rate-Process}.}

\subsection{Demand Aggregation Intervention in New Jersey}
	
In 2014, as part of the New Jersey DRLAP, the state Department of Education planned a coordinated procurement process for broadband internet services, known as the Internet Cooperative Purchasing Initiative (ICPI, also referred to as the Consortium).
Schools that choose to participate in the Consortium 
were asked to submit letters of intent and service order forms for networking and internet access services as part of a \textbf{consolidated Request For Proposals (RFP)} by spring 2014.\footnote{A copy of the RFP can be accessed at \url{https://charliemurry.github.io/ESCNJ_RFP_2014.pdf}.} 
Although the RFP included many services, we focus on dedicated broadband internet services. 
We describe the Internet services in more detail below.
	
	\begin{figure}[t!!]
	\centering
			\caption{Map of New Jersey with Demand Bundling}
		\includegraphics[scale=0.4]{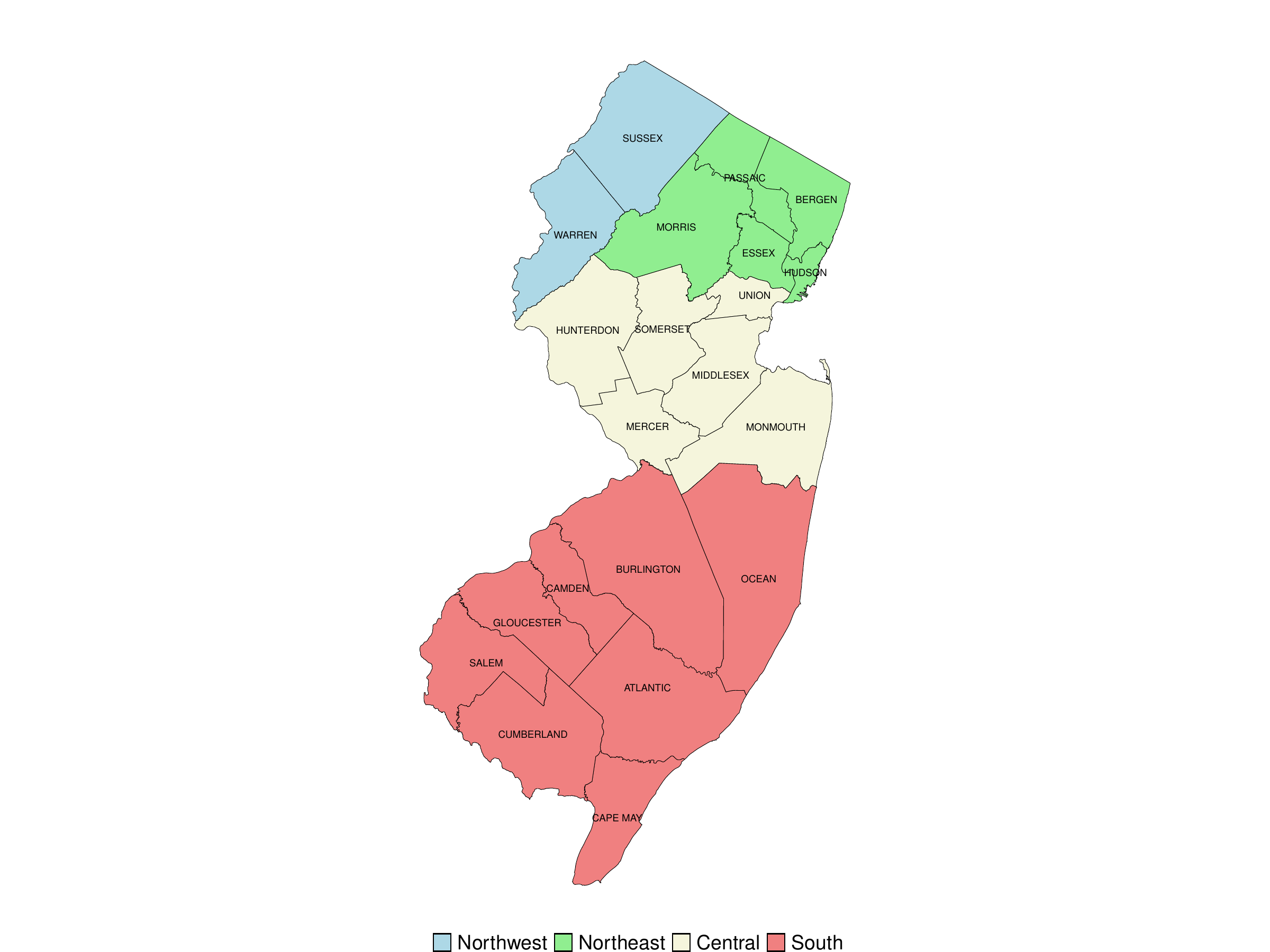}	
		\label{fig:map_regions}
		\begin{figurenotes}
	This figure is a schematic map of the state of New Jersey in the U.S. It contains county boundaries, and each color denotes one of the four regions. 
		\end{figurenotes}
	\end{figure}

Schols were bundled into four groups based on their regions. The regions are illustrated in Figure \ref{fig:map_regions}. After collecting demand information through the RFP, New Jersey conducted a low-bid first-price auction for the bundled school broadband services.  
The structure of the bidding for the Consortium can be best thought of as \emph{pure bundling}, and in this regard, differed from \emph{combinatorial auctions}, as ISPs could not place different bids for a subset of schools. The bids were weighted by qualitative factors such as the ISP's ability to provide service coverage that complied with the technical specifications, the deployment plan, company experience, and service support. 

Some ISPs were household residential providers, and others were large commercial backbone providers.  
Schools also completed a survey about their service level and price as of May 2014, i.e., prior to the introduction of the bundling policy. 
	
The ISP that won the procurement rights for a region guaranteed to deliver the internet to \emph{all} participating schools. The state handled the E-rate paperwork for the participants so that the winning ISP would be the official E-rate provider for these schools if they did not back out. Importantly, schools did not have to accept the terms after the procurement process ended, and ISPs knew this lack of commitment from schools before they bid. However, the RFP also stated that other (non-participating) schools might be interested after the winners were selected. We observed both schools that were initially interested and those that backed out, as well as schools that signed with a winning ISP in 2015, despite not being officially part of the RFP. 

\paragraph{Two products: Category A and Category D} 

In practice, schools procured two types of internet services, the so-called Category A and Category D. Category A internet service is similar to residential retail internet service. A connection is made from the residential building to the ISP network, and the internet service (or payload) is provided through that connection (or transport). This type of product is most appropriate for single-location schools or small school districts with only a few locations because each location would need a separate connection, and no additional services were provided.\footnote{From the Program Results Report (2015): ``Schools in Broadband Group A preferred to receive their internet access as an asymmetrical service delivered on a best-effort basis directly from an ISP. Group A contained 88 schools. Generally, these schools sought low-cost options like cable modems or Verizon FiOS [akin to products typically purchased by residential customers]. Some schools purchased multiple cable modems to serve different schools [each from a different asymmetrical connection points].''} 

Category D is a business-enterprise internet broadband solution. Under the Consortium RFP, the winning ISP was to provide dedicated internet to schools in the Consortium through a regional data hub, which requires high fixed costs. The schools, or school districts, would be connected to the hub through dedicated transport. Large school districts have a private wide area network (WAN) with a connection point that connects to the regional hub. Also, the ISP guarantees internet connectivity (``up-time'') for this product, a typical business enterprise service. A typical purchaser of this product is a medium to large school district that has an existing WAN (or plans to deploy a WAN).\footnote{From the Program Results Report (2015): ``Category D represented dedicated Internet access for schools seeking high capacity, symmetrical Internet access. It was either delivered after being purchased in bulk and distributed through regional WANs (e.g., Affiniti and Lightpath) or distributed directly (e.g., Comcast). Each school requesting service listed its desired amount of Internet access in the RFP.''}$^,$\footnote{The Consortium also solicited services for installing and managing WANs for school districts, but this product is fundamentally different, and we do not study it.}

Both products were offered at the same regional bundle, but the winner in each category could be different. For example, Comcast may have won the Category A contract for the Southern region, but Verizon could win the Category D contract and thus provide a regional hub.

%% file: data.tex
\section{Data Used in the Analysis}

Our main data source is the \cite{NJDRLAP} (ESCNJ, formerly Middlesex Regional Educational Services Commission), which ran surveys to collect information on internet contracts before and after the program. We supplement these data with E-rate data from the FCC through Form 471 \citep{FCC} and Fixed Broadband Deployment data \citep{FCC2}.

\subsection{Data Sources}
The ESCNJ provided us with the data they collected throughout the implementation of the Consortium. 
The information relevant to our analysis includes details of broadband contracts for all public school districts, charter schools, and private schools (including religious schools) that file for the FCC E-rate subsidies in the state. We observe all major contract terms before and after the Consortium, such as the contract price, delivery medium (e.g., fiber, DSL), download and upload bandwidth, servicing ISP, location, and other information about the school district. 
We also observe whether the school chose to participate in the Consortium. 

We define a participant as a school that signed a broadband contract with the winning ISP chosen by the Consortium. 
We treat all other schools as non-participants, even though some have contracts with winning ISPs, for example, because they had an existing multi-year contract with that ISP.
There are 610 schools in our sample. Out of that sample, 327 schools responded to the initial advertisement for the Consortium (some of them responded positively, and some of them responded negatively). Ultimately, 145 schools participated in the Consortium because they signed a contract with the winning ISP through the Consortium. For our baseline analysis, we treat all other schools as part of the control group, although we also perform sensitivity analysis with respect to this definition. 

The program administrator pointed out some reasons for incomplete take-up of the program, including (a) some schools may not have thought the program was beneficial, (b) some schools were in the middle of multi-year contracts, (c) schools typically outsource E-rate procurement to third party consultants who may have different incentives than the schools for joining the Consortium, and (d) this is the first time the program was offered, so schools may not have fully understood the benefits from participating. 

We merge the ESCNJ data with information from the FCC Fixed Broadband Deployment, which includes Census block-level information about which ISPs have active subscribers. Using the FCC data, we construct a measure of the number of active residential and commercial service providers in each school district, which we use as a proxy for competition among ISPs.
For our analysis, we define the bandwidth (or ``speed'') of the internet to be the quoted download maximum in megabits per second (Mbps), and we take the price paid by the school to be the monthly dollars per (download) Mbps. We restrict our analysis to schools with prices below the 95th percentile of the observed price distribution.\footnote{The raw ESCNJ data combine survey data with existing FCC E-rate data. In consultation with the program administrator, we identified numerous observations with extremely high prices that appear anomalous. For example, some schools report prices, \$/Mbps, in the hundreds and thousands. These observations are almost exclusively from 2014, and we exclude them from our analysis.} This leaves us with 1,120 school-year observations across 2014 and 2015 that have complete information on contracts and can be merged with the FCC data.

\subsection{Description of Broadband Contracts}

Schools receive internet through many transport types from ISPs through multi-year contracts, including fiber, coaxial (copper cable), and digital subscriber line (DSL/telephone). Fiber is the highest quality/bandwidth transport medium and is the only way to connect many devices at high speeds with high-quality connections. However, in New Jersey, Comcast is a major ISP (headquartered in Philadelphia) and it distributes the internet through its existing transport network, much of which is coaxial, although Comcast began deploying fiber. In 2014, coaxial connections reached a maximum of 200 Mbps, whereas fiber connections could provide up to 10,000 Mbps of bandwidth per connection.\footnote{The coaxial DOCSIS 3.0 technology that achieved 200 Mbps was introduced in 2006. The DOCSIS 3.1 protocol, which achieved the speed of up to 1,000 Mbps, was introduced in 2013 but not widely adopted in our sample.} We display the presence of ISPs, measured by the number of contracts they sign with schools, across the four regions for 2014 and 2015 in Appendix Table \ref{tab:counts}.

We present descriptive statistics of the contracts in our sample in Table \ref{tab:sumstats}. We separately describe the data for 2014 and 2015, and the participation status (participant or non-participant). Prices are defined as monthly dollars per Mbps per month, and bandwidth is the maximum quoted download Mbps. There is substantial variation in both price and bandwidth across schools, both within and across regions. We display descriptive statistics by geographic region in Appendix Table \ref{tab:sumstats_region}. Prices are typically higher and bandwidth lower in the Northwest region, a rural area of New Jersey. There was also a large overll decrease in prices and an increase in bandwidth from 2014 to 2015 which was part of a general trend of decreasing prices and improved transport networks nationwide during this period.\footnote{As a frame of reference, according to the FCCs ``Urban Rate Survey,'' the average residential connection speed for urban areas of the U.S. in 2014 was 53 Mbps. \url{https://docs.fcc.gov/public/attachments/DA-14-520A3.pdf}}  

\input{tables/tab_sumstats_safe.tex}

In Table \ref{tab:sumstats}, we also display contract terms and connectivity measures broken out by those schools that participated in the Consortium and those that did not. 
Looking at Table \ref{tab:sumstats}, in 2014, participants and non-participants had similar average prices and bandwidth. 
Both groups experienced large price decreases from 2014 to 2015, although participants experienced a larger average decrease. 
Likewise, both groups experienced increases in average bandwidth, although participant increases were much larger. 

We also report a measure of the connectivity of schools to existing networks. We spatially merge the Fixed Broadband Deployment data with the school district's GIS boundaries. We count how many ISPs serve at least one residential building in the school district and report that as the number of active ISPs in a school district (``Number of  ISPs''). Both participants and non-participants average roughly six active ISPs.\footnote{We took this measure in 2014 only; therefore, the table reports this measure twice across both years.} Finally, we also show other relevant connectivity measures in Table \ref{tab:sumstats}. The proportion of schools receiving fiber transport increased from 0.69 to 0.80, while the proportion of schools receiving coaxial decreased from 0.29 to 0.20 between 2014 and 2015.

%% file: tables/tab_sumstats_safe.tex
\begin{table}[t!!]
\small\centering
\caption{\label{tab:sumstats}Summary Statistics}
\small\centering
\begin{threeparttable}
\resizebox{\ifdim\width>\linewidth\linewidth\else\width\fi}{!}{
\begin{tabular}[t]{>{\raggedright\arraybackslash}p{5cm}cccccc}
\toprule
\multicolumn{1}{c}{ } & \multicolumn{3}{c}{Pre Consortium (2014)} & \multicolumn{3}{c}{Post Consortium (2015)} \\
\cmidrule(l{3pt}r{3pt}){2-4} \cmidrule(l{3pt}r{3pt}){5-7}
\textbf{Outcome} & \textbf{Mean} & \textbf{Median} & \textbf{SD} & \textbf{Mean} & \textbf{Median} & \textbf{SD}\\
\midrule
\addlinespace[0.3cm]
\textbf{Price} & \textbf{16.58} & \textbf{10.78} & \textbf{16.31} & \textbf{11.39} & \textbf{6.44} & \textbf{12.41}\\
--Participant & 16.57 & 12.21 & 15.66 & 6.40 & 5.33 & 4.05\\
--Non-participant & 16.58 & 10.25 & 16.53 & 12.95 & 7.70 & 13.67\\
\addlinespace[0.5em]
\textbf{Bandwidth} & \textbf{278.09} & \textbf{100.00} & \textbf{815.38} & \textbf{597.79} & \textbf{150.00} & \textbf{1476.16}\\
--Participant & 288.52 & 100.00 & 388.27 & 1280.76 & 1000.00 & 2424.02\\
--Non-participant & 274.81 & 100.00 & 909.53 & 384.83 & 100.00 & 919.37\\
\addlinespace[0.5em]
\textbf{Number of ISPs} & \textbf{6.05} & \textbf{5.00} & \textbf{2.76} & \textbf{5.96} & \textbf{5.00} & \textbf{2.73}\\
--Participant & 6.18 & 5.00 & 2.65 & 5.90 & 5.00 & 2.68\\
--Non-participant & 6.02 & 5.00 & 2.80 & 5.98 & 5.00 & 2.75\\
\addlinespace[0.5em]
\textbf{Fiber} & \textbf{0.69} & \textbf{--} & \textbf{--} & \textbf{0.80} & \textbf{--} & \textbf{--}\\
--Participant & 0.76 & -- & -- & 0.97 & -- & --\\
--Non-participant & 0.67 & -- & -- & 0.74 & -- & --\\
\addlinespace[0.5em]
\textbf{Coaxial} & \textbf{0.29} & \textbf{--} & \textbf{--} & \textbf{0.20} & \textbf{--} & \textbf{--}\\
--Participant & 0.23 & -- & -- & 0.03 & -- & --\\
--Non-participant & 0.30 & -- & -- & 0.25 & -- & --\\
\addlinespace[0.5em]
\textbf{Other} & \textbf{0.02} & \textbf{--} & \textbf{--} & \textbf{0.00} & \textbf{--} & \textbf{--}\\
--Participant & 0.01 & -- & -- & 0.00 & -- & --\\
--Non-participant & 0.02 & -- & -- & 0.01 & -- & --\\
\textbf{Category D} & \textbf{0.65} & \textbf{--} & \textbf{--} & \textbf{0.67} & \textbf{--} & \textbf{--}\\
--Participant & 0.75 & -- & -- & 0.79 & -- & --\\
--Non-participant & 0.61 & -- & -- & 0.63 & -- & --\\
\bottomrule
\end{tabular}}
\begin{tablenotes}[flushleft]
    \item \footnotesize{\emph{Note}: This table presents summary statistics for our sample of New Jersey schools, disaggregated by year and participation in the ESCNJ Consortium. The sample includes 145 participating schools and 465 non-participating schools.}
\end{tablenotes}
\end{threeparttable}
\end{table}

%% file: analysis.tex

\section{Effects of Demand Bundling}
In this section, we analyze the effect of consortium participation on school broadband prices and bandwidth. 
Recall that we define a participant as a school that responded affirmatively to the initial ESCNJ request for information and signed a broadband contract with the winning ISP chosen by the Consortium.
For our baseline analysis, we classify all other schools as control or non-participants, even though some have contracts with winning ISPs.
Our data lend to an event study strategy to estimate if the Consortium induced better outcomes among participants. 
In particular, we use a DiD strategy to compare the mean outcomes for participating schools to non-participating schools before and after the Consortium. 
Then, using the structure of services provided by the Consortium, we provide evidence that suggests that the effects are primarily because bundling demand mitigates the exposure problem, as opposed to effects through competition. 

However, as we discuss below, there may be a concern that parallel trends may not hold in our setting. 
To assess the robustness of our findings, we follow insights from \cite{manski2018right} and report event study estimates under different degrees of violation of the parallel trends assumption between participants and non-participants.

\subsection{Event Study}

We treat ESCNJ consortium participants as a treated group and all other schools as a control group to estimate the average treatment effect on the treated for price and broadband using the DiD strategy. In particular, we estimate the following regression specification to determine the average treatment on the treated under the DiD assumptions: 
\begin{eqnarray}
Y_{it} &=&  \beta_0 + \beta_1\times \text{Participant}_{it} + \beta^{\texttt{Trend}}\times \text{Post-consortium}_{it} \notag\\&&+ \beta^{\texttt{DiD}} \times (\text{Participant}_{it}\times \text{Post-consortium}_{it}) + X^{\top} \gamma +  \omega_{it}, \label{eq:did}
\end{eqnarray}
where $Y_{it}$ is the outcome variable (price per Mbps or broadband) for school $i$ in year $t\in\{2014, 2015\}$, $\text{Participant}_{it}\in\{0,1\}$ is a binary variable equal to one if $i$ is a participant and zero otherwise, $\text{Post-consortium}_{it}\in\{0,1\}$ is also a binary variable that is equal to one to denote the year when the Consortium was set, and $X$ is a vector of controls. Therefore, in this ``two-by-two" setting, $\text{Post-consortium}_{it}$ is zero for all schools in $t=2014$.

The estimation results from (\ref{eq:did}) are shown in Table \ref{tab:did_main}, with our preferred specifications in columns (3) and (4), which include additional control variables. 
Our parameter of interest is $\beta^{\texttt{DiD}}$, which, under the DiD assumptions, is the effect of consortium participation for participating schools on the outcome variable. The estimates suggest that participation in the Consortium reduced the price of the internet by \$9.37 per Mbps and increased the chosen broadband speed by 308.54 Mbps. Both estimates are statistically significant. 
Thus, under the parallel trends assumption, we conclude that demand bundling caused the price to decrease and the demand for bandwidth to increase.

\input{tables/did_results_safe.tex}

\subsection{Exposure versus Competition}
	There are at least two reasons that bundling could lead to a price drop in our setting. First, ISPs face exposure risk when bidding on schools separately—they may win some contracts but not others, preventing them from fully realizing returns to scale. As our model highlighted in Section \ref{sec:model}, bundling can correct this exposure problem, delivering lower prices when returns to scale are sufficiently large, although there is a countervailing force due to cost heterogeneity (see Figure \ref{fig:bids_comparision}). Second, bundling schools may attract more bidders for the bundle than for separate procurements, and increased competition should lead to lower winning bids (see Table \ref{tab:average_bids}).\footnote{Bundling may also have the opposite effect on the number of bidders. If the bundle sizes are too large, smaller ISPs may be unable or unwilling to serve the entire bundle, as was the case with the central region in our sample (see Table \ref{tab:consortium_bidders}).}

\subsubsection{Evidence of Exposure Effect}	
Although the policy change was not designed to disentangle the exposure effect from the competition effect, we present suggestive evidence that correcting the exposure problem is the dominant reason for the price decline by exploiting the differences in the two internet services offered by the Consortium. 
As discussed in Section \ref{sec:setting}, the New Jersey ICPI procured two types of internet service products.
	
	\begin{description}
		\item[Category A:] A connection (transport) and internet service (payload) to the ISP's existing network. This product was designed for single-location schools and small school districts.
		\item[Category D:] Dedicated transport to a regional hub and internet service (payload) through this hub. This service was designed for medium to large school districts and provided enterprise-level service, including guaranteed uptime.
	\end{description}

The severity of the exposure risk likely differs between Category A and Category D products, driving heterogeneous effects of participation on prices and bandwidth. Category A  involves connecting buildings to existing infrastructure, presenting a relatively modest exposure problem despite some variation in infrastructure proximity. In contrast, Category D service, which requires connecting schools to dedicated regional hubs, faces substantially higher fixed costs, leading to exposure challenges. 
These projects often require new transport infrastructure and hub installations, creating high fixed costs that can only be efficiently amortized across multiple schools.

\input{tables/did_results_catAD_safe.tex}

Table \ref{tab:did_cats} presents separate event study results for each category. 
Columns (5) and (6) show price and bandwidth effects for Category A, and Columns (7) and (8) show results for Category D. 
Consistent with our expectations regarding exposure problems, we find no significant effects for participants in Category A. 
In contrast, Category D participants exhibit effects that are almost identical to our baseline estimates. 
The results suggest that exposure was a major factor in generating price declines due to Consortium participation.

In the remainder of our paper, we focus on Category D only because the benefits of the Consortium are limited to this product, which is also most commonly demanded by schools.

\subsubsection{Evidence of Competition Effect}

Next, we assess the contribution of competition to the observed price reductions. Although we do not have direct measures of competition, the data indicate that schools have an average of 6 ISPs in their location that can serve both residential and commercial customers (Table \ref{tab:sumstats}). So far in our event study analysis (Table \ref{tab:did_cats}), we have used this information to estimate how broadband contract prices vary with the number of ISPs that can provide service, while controlling for the number of ISPs with an active presence in a school district.

As shown in Column (7) of Table \ref{tab:did_cats}, broadband prices decrease by approximately \$0.52 per Mbps for each additional present ISP for Category D services. Given that the main effect of the Consortium is -\$9.17, all else equal, and assuming a linear relationship between competition and prices, we would require approximately 17 ISPs for competition to have the same effect on prices. This magnitude is unlikely for two reasons. First, it would require more ISPs than we observe in our sample. Second, the assumption of linearity likely overstates the competitive effect, as the impact of additional competitors typically exhibits diminishing returns \citep{Watt2024}.

However, to more formally investigate the role of competition in determining the effect of Consortium prices, we use data on the number of bidders involved in the Consortium procurement in the event study design. 
Table \ref{tab:consortium_bidders} presents the number of Consortium bidders by region for Category D services. The Consortium required ISPs to bid for the entire region (pure bundling), but some submitted partial bids covering only some schools within a region. 
We distinguish between the number of complete bidders (only those who agreed to serve all schools in the region) and the total number of bidders (including those who did not agree to serve the entire region). The data reveal limited competition in terms of complete bids: most regions received two complete bids, with the Central region receiving zero complete bids, necessitating that the Consortium program administrator split the contract award between two ISPs. 

\begin{table}[t!!]
\centering
\caption{Consortium Bidding Competition by Region}
\label{tab:consortium_bidders}
\begin{threeparttable}
\begin{tabular}{lcccc}
\toprule
 & Listed & & & \\
Region & Schools & Winner(s) & Complete Bidders & Total Bidders \\
\midrule 
Central & 95 & Comcast/Lightpath & 0 & 11 \\
Southern & 102 & Comcast & 2 & 10 \\
Northeast & 114 & Sunesys & 2 & 11 \\
Northwest & 21 & Sunesys & 2 & 6 \\
\bottomrule
\end{tabular}
\begin{tablenotes}[flushleft]
\item {\footnotesize\textit{Note:} This table shows bidding competition for Category D services in Consortium auctions. Complete bidders are those bidders who agreed to serve all participating schools in the region, and total bidders include those that also submit partial coverage bids. The Central region was split into two parts after the completion of the procurement process due to the lack of complete bids.}
\end{tablenotes}
\end{threeparttable}
\end{table}

We construct two alternative proxies of competition to measure the competitive effects of the Consortium. The first uses complete bidders as the relevant competition measure, representing a lower bound on competition because only these ISPs could win contracts. The second uses total bidders, representing an upper bound on competition. For participating schools, we replaced the ISP count used in Table \ref{tab:did_cats} for 2015 with these numbers. 

For example, let us say that Westhampton School District had five active residential ISPs in its geographic area in 2014, and it joined the Consortium in the Southern region in 2015. The measure of competition we would assign to Westhampton in 2014 is five ISPs. In 2015, we would assign either two ISPs (based on the number of complete bids) or ten ISPs (based on the total number of bids). If Westhampton had not joined the Consortium, we would assign it five ISPs as a measure of competition in both years.

Table \ref{tab:competition_bounds} presents the DiD estimates using these alternative competition measures. Column (9) uses total bidders, yielding a Consortium effect of -\$6.61. Column (10) uses complete bidders, producing a Consortium effect of -\$12.32. Our baseline estimate of the effect of the Consortium -\$9.17 falls between these bounds.
The coefficient on the competition measure is similar for both cases (-0.59 and -0.63). 
The results reveal that controlling for competition has an impact on the findings. 
Whether the Consortium induced more competition on average (represented by total bids) or less competition (represented by complete bids) is unknown from the available data, but controlling for competition changes the estimates of the price effect of the Consortium. 

\input{tables/did_competition_bound_catD_safe.tex}

However, these estimates suggest that competition cannot fully explain the observed price effects. Even under the most generous assumptions about competitive pressure, the Consortium lowers prices by \$6.61/Mbps. The similarity of competition coefficients across specifications and our baseline estimates provides confidence that exposure effects from demand bundling represent the primary mechanism driving price reductions.

\subsection{Sensitivity Analysis}

\subsubsection{Violations of Parallel Trends}
The Consortium was marketed to all schools and school districts in New Jersey participating in \emph{E-rate}. 
Their participation was voluntary, and as discussed earlier, not all schools joined the Consortium. 

As such, participants may have different baseline trends in prices and bandwidth than non-participants. 
First, participating schools may join the Consortium because they have poor prospects for gaining cheaper and faster internet through their current procurement process. 
This ``negative'' selection into the program would imply that our DiD estimate understates the true effect. 
Second, schools that think they have the most to gain are also in areas that would naturally have steeper gains. They could be in areas that are ``catching up'' in broadband availability. 
This ``positive'' selection into the program would imply that our DiD estimate overstates the true effect. 

We explore how our estimates would change if the parallel trend assumption were violated.
To that end, we take an approach suggested by \cite{manski2018right} and determine the DiD estimate if the trend for the treatment groups were $g\in\mathbb{R}_+$ times the trend for the control groups, such that $g=1$ denotes the baseline with parallel trends.  
With a slight abuse of notations, we consider violations of the parallel trends assumption of the form
\begin{eqnarray}
	\hat{\beta}^{\texttt{Robust}} &=& \hat{\beta}^{\texttt{DiD}} + \Big(\hat{\beta}^{\texttt{Trend,0}}-\hat{\beta}^{\texttt{Trend,1}}\Big) =
	\hat{\beta}^{\texttt{DiD}} + \Big(\hat{\beta}^{\texttt{Trend,0}}-g\times \hat{\beta}^{\texttt{Trend,0}}\Big),\label{eq:bounds}
\end{eqnarray}
where $\hat{\beta}^{\texttt{Trend,0}}$ is the trend for the control group, $\hat{\beta}^{\texttt{Trend,1}}$ is the (unidentifiable) trend for the participants. 
For the second equality we imposed that $\hat{\beta}^{\texttt{Trend,1}}=g\times \hat{\beta}^{\texttt{Trend,0}}$ with $g\geq0$ is  the degree of violation of the parallel trend. Note that this setup nests our baseline specification with parallel trend assumption ($\hat{\beta}^{\texttt{Trend,0}}=\hat{\beta}^{\texttt{Trend,1}}$) as a special case, i.e., when $g=1$, $\hat{\beta}^{\texttt{Robust}}=\hat{\beta}^{\texttt{DiD}}$, identifying the causal effects of the Consortium. We ask, ``What if the participants had a $g$ times the trend of the control schools?"

\begin{figure}[t!!]
\centering
\caption{Estimates Accounting for Violations in Parallel Trends}
	\includegraphics[width=0.48\textwidth]{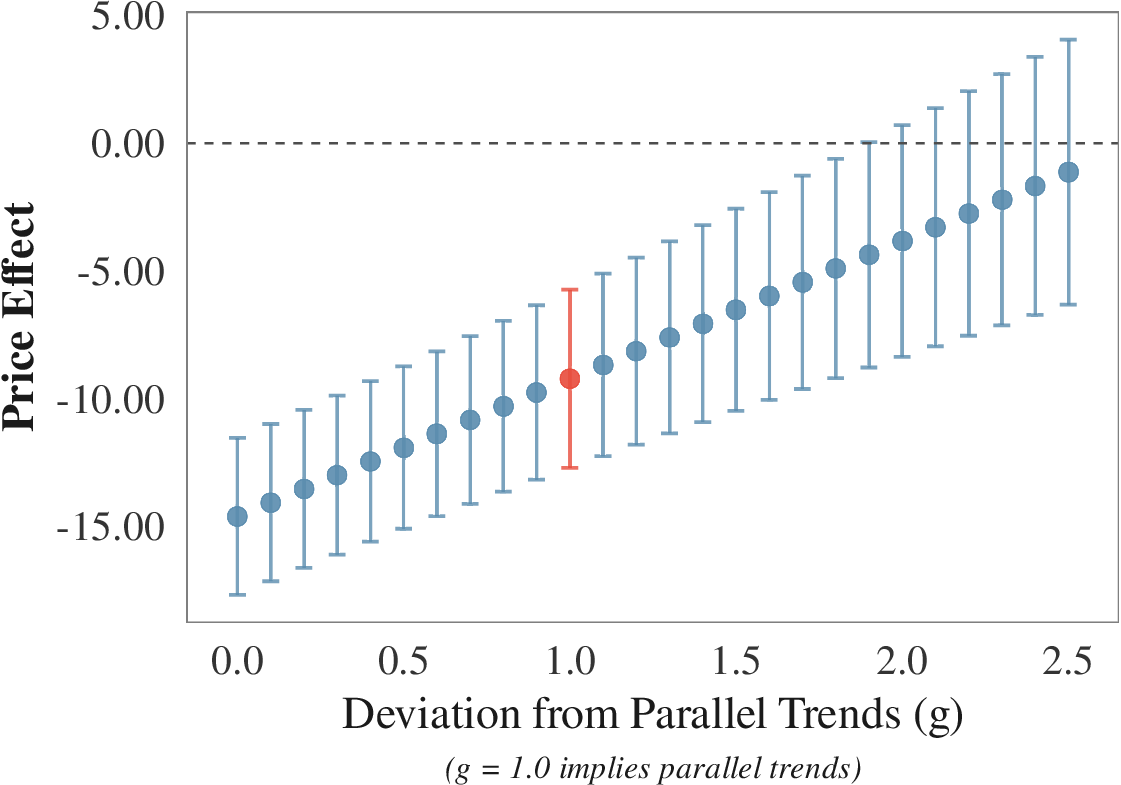}~
	\includegraphics[width=0.48\textwidth]{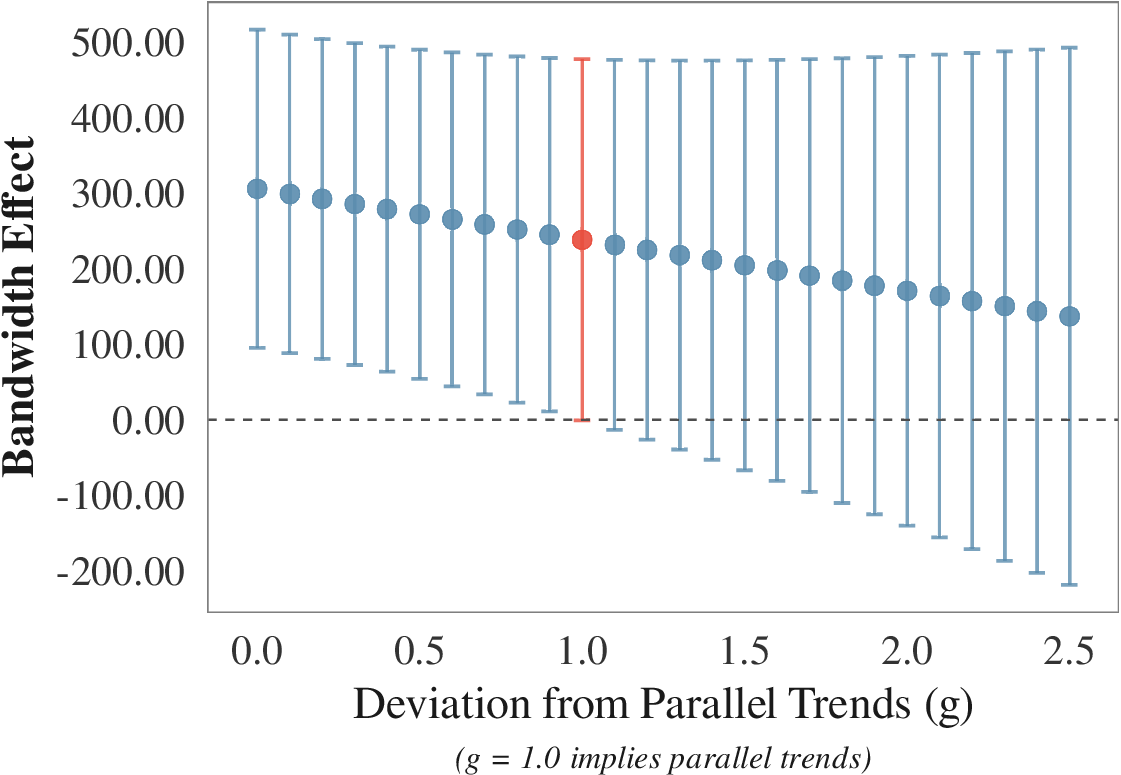}
	\label{fig:bounds}
	\begin{figurenotes}
		This figure displays robust coefficient estimates ($\hat{\beta}^{\texttt{Robust}}$) from equation (\ref{eq:bounds}) for both price (left panel) and broadband demand (right panel). Each panel shows the estimates and their 95\% confidence intervals across for $g\in \{0, 0.10, \cdots, 1, \cdots, 2.5\},$  which represents the magnitude of potential parallel trends violations.
		\end{figurenotes}
\end{figure}

In Figure \ref{fig:bounds}, we present the estimates of $\hat{\beta}^{\texttt{Robust}}$ (represted as a ``dot") from (\ref{eq:bounds}) using the estimates for the Category D product in Table \ref{tab:did_cats} for  $g\in \{0, 0.10, \cdots, 1, \cdots, 2.5\}$. 
For instance, to estimate $\hat{\beta}^{\texttt{Robust}}$ for price, we use $\hat{\beta}^{\texttt{Trend,0}}=-5.53$ and $\hat{\beta}^{\texttt{DiD}}=-9.17$ (from Table \ref{tab:did_cats} Column (7), coefficients Post-consortium and Particiapant $\times$ Post-consortium, respectively). 
We use the standard errors and covariance from Table \ref{tab:did_cats} to determine the 95\% confidence intervals.\footnote{\cite{manski2018right} have a long panel and use past data to inform the level of violation ($g$). Furthermore,  \cite{rr_honest2023} propose a uniformly valid inference procedure when the parallel trends assumption is violated. In our ``two-by-two" setup, we do not have past data that can inform $g$, underpinning our choice of a fixed $g$.
Furthermore, our confidence intervals do not consider sampling variability that may affect non-participant trends, nor do we adjust for multiple testing.} 

The red dot is our baseline estimate with $g=1$. 
Next, consider the price effect under the extreme case of $g = 2.5$. That is equivalent to saying that the participant trend would have been twice and a half as steep as the non-participant trend, i.e., price would have decreased by  $2.5 \times -\$5.53 = -\$13.82$ for the non-participants, so that the treatment effect would be $-\$9.17+ (-\$5.53 + \$13.82) = -\$0.88$. In other words, the treatment effect would disappear if the price trend for participating schools was two and a half times as steep as that of non-participants. 
The estimates suggest that the treatment effect would be completely erased if participant price trends were slightly more than two and a half times as steep (i.e., decreasing) as those of non-participants. Taking into account confidence intervals, the statistical significance of the robust estimate disappears if the participant price trend is approximately 1.7 times that of the non-participant price trend.

The treatment effect on bandwidth appears less sensitive to the assumption of parallel trends. 
In other words, the treatment effects are less steep than those for the prices, so that it would take the difference in trends more than two and a half times to ``wash out" our estimated effect. 
However, the estimates are less precise, and the statistical significance disappears if the trend for participants is steeper than that for non-participants.

\subsubsection{Definition of Control Group}
Although the treated and control groups look fairly similar in the pre-treatment period (see Table \ref{tab:sumstats}), we assess the robustness of the definition of the control group.
Some schools may not be actively choosing a new ISP in 2014 when the Consortium was announced because they may have signed a previous multi-year contract. If so, those schools may not be the appropriate control group. As a robustness exercise, we exclude schools with the same contract (i.e., the same ISP, bandwidth, and price) in 2014 and 2015 that we previously identified as being in the control group. 
This process eliminates 192 school-year observations from the analysis. The results with this alternative sample are in Columns (11) and (14) of Table \ref{tab:did_robust}, and they are similar to the baseline estimates.

Additionally, some schools that did not participate in the Consortium may still be impacted because they can contract with the winning ISP at a lower (winning bid) price. 
Again, these schools may not be the appropriate control schools. 
For the next robustness specification, we exclude those schools that did not participate in the Consortium but signed 2015 contracts with the winning ISP in their region. 
There were 236 such school-year observations. The results with this alternative sample are in Columns (12) and (15) of Table \ref{tab:did_robust}. Excluding those non-participants who contract with winners yields a similar effect of bundling as our primary estimates.

\input{tables/did_robustness_64_safe.tex}

\subsection{Selection}
Our setting permits us to consider the potential impact of joining the Consortium on those who \emph{did not} participate. 
This counterfactual analysis is a way to investigate whether selection drives our results. 
Specifically, we examine the potential savings for non-participating schools had they individually joined the Consortium in 2015. 
In other words, we compare the available Consortium contracts with those of non-participants and determine whether they would have been better off joining the Consortium. 

We focus specifically on schools that neither expressed interest in the program nor held contracts with the winning ISP in their region, allowing us to keep our analysis clean.
As before, we continue to restrict the analysis to schools purchasing a Category D broadband product, that is, high-speed enterprise-level service. 
Sixty-nine schools in our sample did not participate in the Consortium, contracted with a non-winning ISP in 2015, and purchased bandwidth available from their region's winning Consortium ISP. 

\begin{figure}[t!!]
	\centering
		\caption{Price Difference Between 2015 Contract and Winning Bid for Control Schools}
		\includegraphics[scale=0.4]{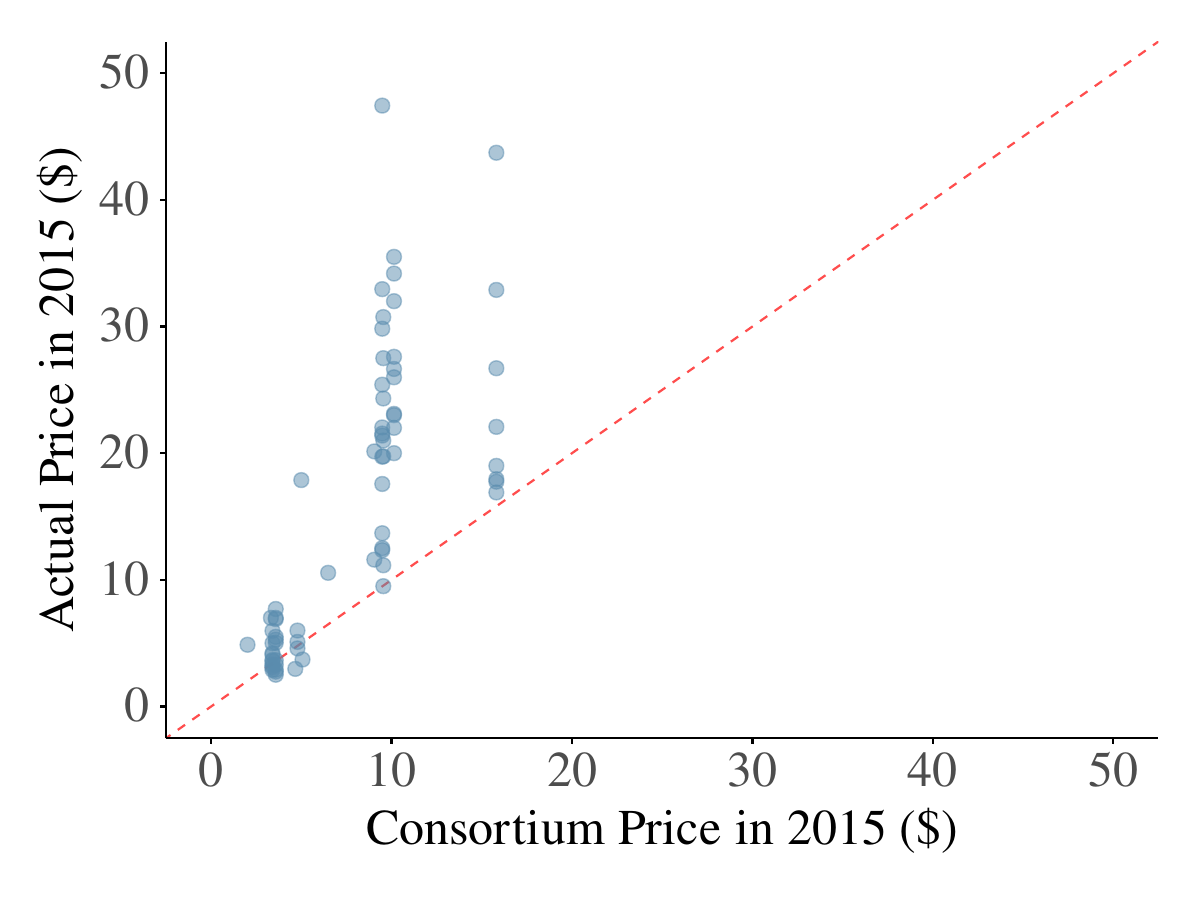}
		\label{fig:untreated}
		\begin{figurenotes}
		Scatter plot of the 2015 actual prices and the Consortium prices for the schools' chosen bandwidths. Represents all schools that did not express interest in the program and did not sign a contract with a Consortium winner.
		\end{figurenotes}
	\end{figure}

Figure \ref{fig:untreated} compares actual 2015 prices with Consortium prices for these 69 schools. We find that 57 schools paid higher prices than they would have under the winning ISP's Consortium bid. 
Among these schools, the average potential savings from joining the Consortium would have been \$9.47 per Mbps, similar to the baseline estimate of savings of \$9.17 per Mbps (Table \ref{tab:did_cats}). 
For the 12 schools that would not benefit from switching, the average difference in their price and the Consortium's is \$0.61.

To formalize this analysis, we re-estimate the DiD model using only these 69 schools as the control group while maintaining the same treated group as in Table \ref{tab:did_cats}. The results, presented in columns (13) and (16) of Table \ref{tab:did_robust}, demonstrate that the estimated treatment effects remain qualitatively similar to our baseline findings.

The similarity between potential savings for non-participants and realized savings for participants suggests that selection into treatment is not driving our main results -- schools that chose not to participate appear to have forgone similar benefits to those realized by participants. This finding indicates that factors other than heterogeneous potential gains likely drove participation decisions. 
Future research could explore these participation frictions and the program's dynamic implications. However, such an analysis would require additional data from subsequent years and program waves.

\subsection{SUTVA}

An additional identification concern involves the program's effect on untreated schools because broadband prices are determined in equilibrium. 
SUTVA requires the absence of spillover effects, which may be violated in our setting if a school's participation in the Consortium influences equilibrium bids across the region.

We identify two specific concerns. First, control schools may receive lower prices in 2015 if the winning ISP adjusts its pricing strategy based on new Consortium contracts. 
However, this would only attenuate our estimates of the price difference between treated and control schools, suggesting that, if anything, our base estimates are conservative. 
Second, the program's infrastructure build-out could lower costs for the control group. 
This concern is less likely to affect our estimates because we use only one year before and after the Consortium's establishment, and it takes several years before infrastructure build-out effects materialize.

%% file: tables/did_results_safe.tex
\begin{table}[t!!]
	\centering
	\caption{Estimated Effect of Demand Bundling}\label{tab:did_main}
		\begin{adjustbox}{scale=0.9}
		\begin{threeparttable}
	\begin{tabular}{@{\extracolsep{5pt}}lcccc}
\\[-1.8ex]\toprule
\\[-1.8ex] & Price & Bandwidth & Price & Bandwidth\\
\\[-1.8ex] & (1) & (2) & (3) & (4)\\
\hline \\[-1.8ex]
Non-participant & 16.579$^{***}$ & 306.128$^{***}$ & 21.466$^{***}$ & 62.662 \\
& (0.720) & (45.656) & (4.256) & (285.838) \\
& & & & \\
Participant & 16.575$^{***}$ & 311.059$^{***}$ & 22.005$^{***}$ & $-$20.365 \\
& (1.284) & (81.420) & (4.400) & (296.232) \\
& & & & \\
Post-Consortium & $-$3.625$^{***}$ & 99.713 & $-$4.363$^{***}$ & 79.520 \\
& (0.975) & (61.836) & (0.858) & (58.932) \\
& & & & \\
Participant x Post-Consortium & $-$6.553$^{***}$ & 307.263$^{**}$ & $-$9.376$^{***}$ & 308.544$^{**}$ \\
& (1.997) & (126.612) & (1.780) & (122.094) \\
& & & & \\
Number of ISPs &  &  & $-$0.364$^{**}$ &  \\
&  &  & (0.156) &  \\
& & & & \\
\midrule
School Type &  &  & $\checkmark$ & $\checkmark$ \\
ISP &  &  & $\checkmark$ & $\checkmark$ \\
Region &  &  & $\checkmark$ & $\checkmark$ \\
Service Type &  &  & $\checkmark$ & $\checkmark$ \\
\\[-1.8ex]
Observations & 1120 & 1120 & 1120 & 1120 \\
\bottomrule
\end{tabular}
\begin{tablenotes}[flushleft]
\item \footnotesize \emph{Note:} This table reports difference-in-differences estimates from equation (\ref{eq:did}) under varying specifications. The analysis is structured as follows: Columns (1) and (2) present baseline estimates without control variables, and Columns (3) and (4) present estimates with a full set of control variables.\\ $^{**}$p$<$0.05; $^{***}$p$<$0.01.
\end{tablenotes}
\end{threeparttable}
\end{adjustbox}
\end{table}

%% file: tables/did_results_catAD_safe.tex
\begin{table}[t!!]
	\centering
	\caption{Effect of Demand Bundling for Different Internet Products}\label{tab:did_cats}
		\begin{adjustbox}{scale=0.9}
		\begin{threeparttable}
	\begin{tabular}{@{\extracolsep{5pt}}lcccc}
\\[-1.8ex]\toprule
 & \multicolumn{2}{c}{Category A} & \multicolumn{2}{c}{Category D} \\
\cline{2-3} \cline{4-5}
\\[-1.8ex] & Price & Bandwidth & Price & Bandwidth\\
\\[-1.8ex] & (5) & (6) & (7) & (8)\\
\hline \\[-1.8ex]
Non-participant & 9.201$^{**}$ & 182.285 & 39.176$^{***}$ & $-$134.346 \\
& (3.705) & (129.567) & (7.033) & (530.623) \\
& & & & \\
Participant & 7.085$^{*}$ & 198.240 & 39.342$^{***}$ & $-$261.340 \\
& (3.787) & (132.690) & (7.260) & (548.890) \\
& & & & \\
Post-Consortium & $-$1.143 & 14.077 & $-$5.539$^{***}$ & 114.837 \\
& (0.758) & (27.468) & (1.200) & (91.864) \\
& & & & \\
Participant x Post-Consortium & $-$0.440 & 87.683 & $-$9.174$^{***}$ & 380.062$^{**}$ \\
& (2.096) & (75.976) & (2.287) & (174.839) \\
& & & & \\
Number of ISPs & $-$0.154 &  & $-$0.521$^{**}$ &  \\
& (0.150) &  & (0.209) &  \\
& & & & \\
\midrule
Observations & 382 & 382 & 738 & 738 \\
\bottomrule
\end{tabular}
\begin{tablenotes}[flushleft]
\item \footnotesize \emph{Note:} This table reports difference-in-differences estimates from equation (\ref{eq:did}) for the two products separately: Category A and Category D (explained in text). All the specifications include school type,
ISP, and region and service type fixed effects. The analysis is structured as follows: Columns (5) and (7) present price as dependent variable for Category A and D, respectively. Columns (6) and (8) present bandwidth as a dependent variable for Category A and D, respectively.\\ $^{**}$p$<$0.05; $^{***}$p$<$0.01.
\end{tablenotes}
\end{threeparttable}
\end{adjustbox}
\end{table}

%% file: tables/did_competition_bound_catD_safe.tex
\begin{table}[t!!]
	\centering
	\caption{Price Effects Controlling for Consortium Competition }\label{tab:competition_bounds}
		\begin{adjustbox}{scale=0.9}
		\begin{threeparttable}
	\begin{tabular}{@{\extracolsep{5pt}}lccc}
\\[-1.8ex]\toprule
\\[-1.8ex] & Price & Price\\
\\[-1.8ex] & (9) & (10)\\
\hline \\[-1.8ex]
Participant x Post-Consortium & $-$6.614$^{***}$ & $-$12.325$^{***}$ \\
& (2.456) & (2.573) \\
& & &  \\
Potential Bidders & $-$0.590$^{**}$ & $-$0.634$^{***}$ \\
& (0.229) & (0.223) \\
& & &  \\
\midrule
Observations & 738 & 738 \\
\bottomrule
\end{tabular}
\begin{tablenotes}[flushleft]
\item \footnotesize \emph{Note:} This table presents DiD estimates of price effects for Category D services using alternative measures of competition. All the specifications include school type, ISP, region, and service type fixed effects. Column (9) uses total bidders (including partial coverage bids) as the competition measure. Column (10) uses only complete bidders who agreed to serve entire regions. Standard errors in parentheses.\\ $^{**}$p$<$0.05; $^{***}$p$<$0.01.
\end{tablenotes}
\end{threeparttable}
\end{adjustbox}
\end{table}

%% file: tables/did_robustness_64_safe.tex
\begin{table}[t!!]
	\centering
	\caption{Robustness Checks: Alternative Sample Specifications}\label{tab:did_robust}
		\begin{adjustbox}{scale=0.9}
		\begin{threeparttable}
	\begin{tabular}{@{\extracolsep{5pt}}lcccccc}
\\[-1.8ex]\toprule
\\[-1.8ex] & Price & Price & Price & Bandwidth & Bandwidth & Bandwidth\\
\\[-1.8ex] &  (11) & (12) & (13) & (14) & (15) &(16)\\
\hline \\[-1.8ex]
Non-participant & 64.922$^{***}$ & 65.619$^{***}$ & 22.978$^{***}$ & $-$263.547 & $-$279.708 & 209.640 \\
& (13.723) & (12.693) & (2.607) & (904.496) & (1050.367) & (152.778) \\
& & & & \\
Participant & 64.087$^{***}$ & 64.780$^{***}$ & 21.376$^{***}$ & $-$358.198 & $-$330.660 & 330.907 \\
& (13.914) & (13.058) & (3.286) & (917.379) & (1081.009) & (210.241) \\
& & & & \\
Post-Consortium & $-$6.558$^{***}$ & $-$6.823$^{***}$ & $-$7.614$^{***}$ & 148.107 & 105.735 & 177.940 \\
& (1.476) & (1.538) & (2.024) & (97.431) & (127.417) & (144.824) \\
& & & & \\
Participant  & $-$7.831$^{***}$ & $-$7.406$^{***}$ & $-$6.696$^{**}$ & 364.674$^{**}$ & 397.998$^{**}$ & 320.186$^{*}$ \\
x Post-Consortium& (2.546) & (2.422) & (2.647) & (168.103) & (200.544) & (189.389) \\
& & & & \\
Number of ISPs & $-$0.466$^{*}$ & $-$0.354 & $-$0.331 &  &  &  \\
& (0.248) & (0.245) & (0.269) &  &  &  \\
& & & & \\
\midrule
Observations & 546 & 502 & 336 & 546 & 502 & 336 \\
\bottomrule
\end{tabular}
\begin{tablenotes}[flushleft]
\item \footnotesize \emph{Note:} This table presents robustness checks using alternative sample specifications. Columns (11) and (14) restrict the sample to active participants only. Columns (12) and (15) exclude impacted schools from the analysis. Columns (13) and (16) include only the schools defined in Section 5.3.3 as controls. All specifications include the full set of control variables and include only Category D schools.\\ $^{**}$p$<$0.05; $^{***}$p$<$0.01.
\end{tablenotes}
\end{threeparttable}
\end{adjustbox}
\end{table}

%% file: welfare.tex

\section{Expenditures and Welfare}

In this section, we measure the effect of the Consortium on the schools' expenditures and welfare. First, we use the DiD estimates from Category D participants to determine the savings from participating in the Consortium. Second, we use observed prices and broadband choices before and after the Consortium to determine the bounds for the change in school welfare for participating and non-participating schools. Then, using the DiD estimates for different degrees of parallel trend violation, we isolate the change in welfare from participating in the Consortium. 
Our main finding is that savings are large relative to the total E-rate subsidy, and the program improves the welfare of the participating schools.

\subsection{Expenditures}

In our 2014 sample, schools that participated in the Consortium and purchased Category D services spent a total of approximately \$4.78 million on internet services, of which \$2.47 million was reimbursed by the \emph{E-rate} program. 
Next, we determine the savings that accrue to the schools from the Consortium and compare them to this \emph{E-rate} subsidy.

As shown in the last section, the Consortium lowered the prices by an average of \$9.17 per Mbps, and increased the chosen bandwidth by 380.06 Mbps. Using these estimates, we determine two savings measures, the Paasche and Laspeyres price indices for participants, which represent lower and upper bounds on savings, depending on the assumption of the chosen bandwidth absent the new prices.

First, we determine the savings by imposing that each school saves $\beta^{\texttt{DiD}}$ per Mbps, with each school's demand held at the 2014 level ($Q_{0}$). 
This exercise determines the lower bound for savings because it keeps the demand fixed at the 2014 level. 
Second, we determine the savings by allowing the broadband demand to increase by the treatment effect of bandwidth, i.e., $Q_{1} = Q_{0} + 380.06$ Mbps, which gives us the upper bound. 
We also compute the effective E-rate subsidy for the schools to benchmark these savings. 

\begin{table}[t!!]
	\centering
	\caption{Bounds on the Expenditure Savings\label{tab:expenditure_bounds}}
	\begin{threeparttable}
\begin{tabular}{p{4cm}cc}
	\toprule
  & Expression & Amount \\
  \midrule
  Lower Bound & 
  $- \hat{\beta}^{\text{price}} \times Q_{0} \times (1-\rho)$ &
  \$1,618,269\\[0.5em]
  Upper Bound & 
  $- \hat{\beta}^{\text{price}}(Q_{0} + \hat{\beta}^{\text{mbps}})\times(1-\rho)$ &
  \$3,482,281 \\[0.5em]
  E-rate Subsidy &
  $Q_{0} \times P_{0} \times \rho$&
  \$2,474,609\\
  \bottomrule
\end{tabular}
\begin{tablenotes}[flushleft]
\item \footnotesize \emph{Note:} $\hat{\beta}^{\text{price}}$ and $\hat{\beta}^{\text{mbps}}$ are the estimates in Table \ref{tab:did_cats} Columns (7) and (8), $P_{0}$ is the price paid in 2014 and $\rho$ is the E-rate subsidy rate. 
\end{tablenotes}
\end{threeparttable}
\end{table}

In Table \ref{tab:expenditure_bounds}, we present the savings and the total E-rate subsidy for all participating schools, aggregated for the entire year. 
We find that the savings range between  \$1.61 million and \$3.48 million. 
These savings translate into 65\% to 140\% of the \$2.47 million total E-rate subsidy that the FCC paid to participating schools in New Jersey. 
Thus, the demand bundling program can achieve similar cost savings to those of the schools or greater bandwidth-adjusted savings at no additional cost to taxpayers. 

\begin{figure}[th!!]
\centering
\caption{Bounds on Savings under the Violation in Parallel Trends}
	\includegraphics[width=0.65\textwidth]{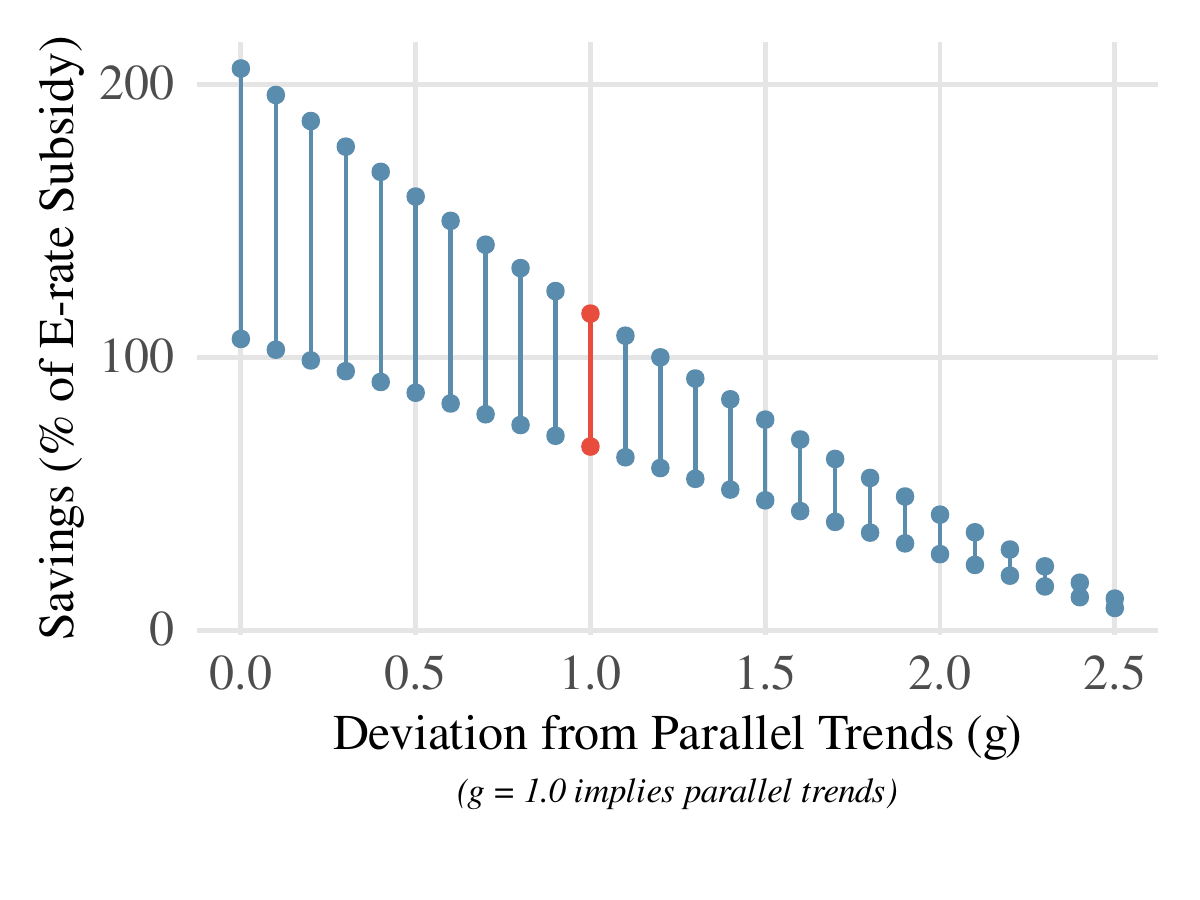}
	\label{fig:bounds_expenditure}
	\begin{figurenotes}
	This figure illustrates the total cost savings achieved by participating schools in 2015, expressed as a percentage of the total E-rate subsidy (\$2.69 million). Each bar represents the upper and lower bounds for savings determined using the formulae in Table \ref{tab:expenditure_bounds} (in columns marked ``Expression"), with estimates based on $\beta^{\texttt{Robust}}$ from equation (\ref{eq:bounds}) across a range of $g\in\{0, 5, \cdots, 1,\cdots, 1.5 \cdots, 2.5\}$ of violation of parallel trends assumption.
		\end{figurenotes}
\end{figure}

Similar to the robustness exercise in Figure \ref{fig:bounds}, in  Figure \ref{fig:bounds_expenditure}, we present the bounds on the savings, expressed as a percentage of \$2.47 million--the total E-rate subsidy paid to participating schools. As we can see, the bounds are strictly positive for violations of parallel trends less than a factor of 2.5.
This evidence suggests that a policymaker deciding between implementing the current E-rate subsidy design and the ESCNJ Consortium design may prefer the latter because the latter achieves similar savings at no direct taxpayer cost.

\subsection{Welfare of Schools}

Measuring the total savings conflates lower prices and higher purchased bandwidth. 
In other words, a school's total cost could be unchanged, but it could be better off because it pays the same amount for greater bandwidth. 
A better measure of the Consortium's effect would be the change in welfare resulting from lower prices and greater bandwidth. 

In this section, we quantify the change in welfare due to the Consortium. 
Let ${\mathcal D}_i(\cdot):\mathbb{R}_+\rightarrow\mathbb{R}_+$ be the demand function for school $i$. For a school $s$, we observe the 
prices and broadband choices pairs (net of the E-rate subsidies) before the Consortium, $(P_{i0}, Q_{i0})$, and after the Consortium $(P_{i1}, Q_{i1})$. We want to determine the change in welfare, $\Delta W_i$, which is the area under the demand curve between two prices:
$$
\Delta W_i = \int_{P_{i1}}^{P_{i0}} {\mathcal D}_i(\xi) d\xi.
$$
Instead of estimating the demand function for broadband for schools, which is difficult given our data, we rely on insights from \cite{kang2022robust} and determine the bounds for $\Delta W_i$.
For simplicity, we assume that the demand function is log-concave, i.e., $\frac{{\mathcal D}_i'(P)}{{\mathcal D}_i(P)}$ is decreasing in $P$ for all schools. \cite{kang2022robust} show that the change in welfare can be bounded as  
\begin{eqnarray}
Q_{i0}(P_{i0}-P_{i1}) \leq \Delta W_i \leq \frac{(P_{i0}-P_{i1})\times (Q_{i1}-Q_{i0})}{\log\left(\frac{Q_{i1}}{Q_{i0}}\right)},\label{eq:cs_bounds}
\end{eqnarray}
where this bound is sharp.\footnote{Log-concavity provides a simple and tractable demand specification that yields intuitive welfare bounds in \cite{kang2022robust}. 
See \cite{CaplinNalebuff1991} for more on log-concave demand. While our specific welfare bounds depend on this assumption, we expect our core finding—that welfare increases due to lower prices and higher quantities—to hold across reasonable demand specifications. See \cite{kang2022robust} for bounds under alternative demand assumptions.} Note that both the lower and upper bounds depend on the data. 
However, before determining these bounds, we select our sample appropriately.

\begin{figure}[t!!]
\caption{Bounds for Change in School Welfare}\label{fig:cs_bounds}	
\begin{subfigure}{0.48\linewidth}
	\includegraphics[width=1.05\textwidth]{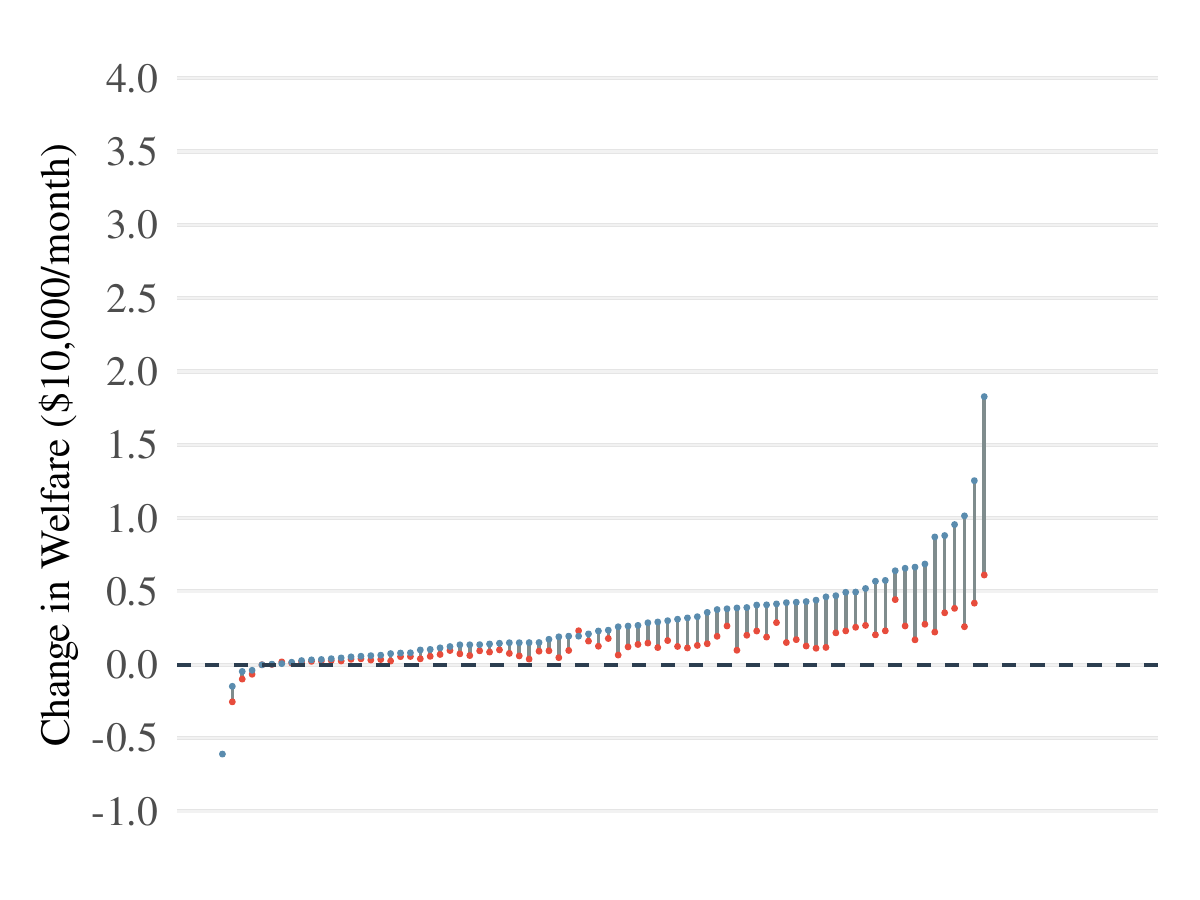}
	\caption{Participants}
\end{subfigure}
	\begin{subfigure}{0.48\linewidth}
	\includegraphics[width=1.05\textwidth]{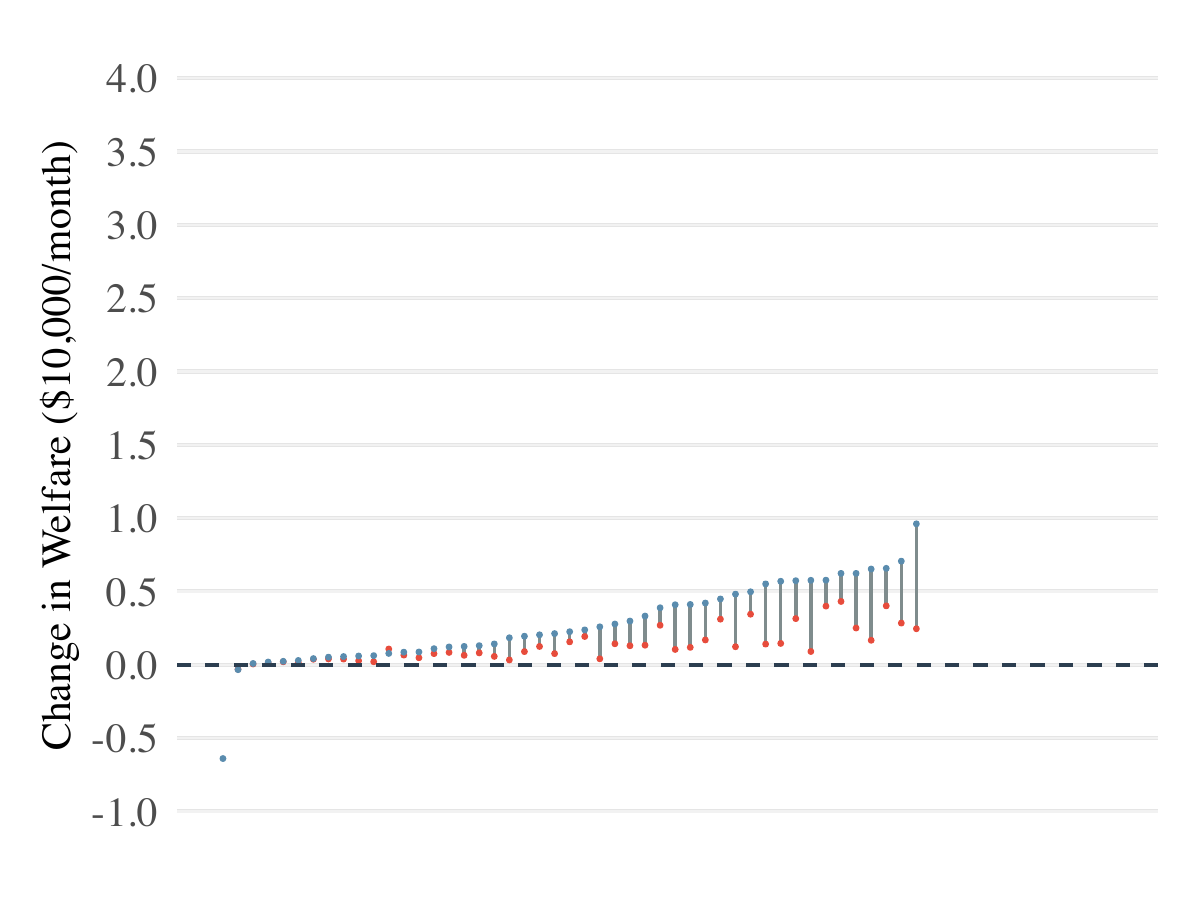}
	\caption{Non-participants}
	\end{subfigure}
	\begin{figurenotes}
	This figure shows the changes in school welfare, measured in \$10,000 increments per month. Blue dots represent the upper bounds, while the lower bounds are shown in red dots, as calculated using equation (\ref{eq:cs_bounds}). The left panel displays results for 117 participating schools, while the right panel shows results for 109 non-participating schools.
		\end{figurenotes}
\end{figure}

Of the 424 schools that are interested in Cat D service, we observe 313 schools in both years. 
We discard schools that upgrade service, for example, from DSL to fiber, as we expect the demand function for different services to be different, and fiber availability may change across the two years. We also discard schools that have choices that are not consistent with downward-sloping demand. 

In Figure \ref{fig:cs_bounds}, we separately present these bounds for participating and non-participating schools. We use observed prices for both periods to compare the welfare across the two groups. 
In both cases, most schools have a positive lower bound, likely because prices decreased and bandwidth increased for all schools between 2014 and 2015, due to the underlying common trends.

The welfare bounds presented in Figure \ref{fig:cs_bounds} for participating schools reflect the total effect, including the effects of trends and product types, as they are computed using observed prices and quantities. To isolate the specific impact of bundling on welfare changes for the participating schools relative to the non-participating schools, we conduct a counterfactual analysis using our estimated price and quantity coefficients to determine the new $P_1=P_0 + \beta^{\text{price}}$ and $Q_1=Q_0 + \beta^{\text{mbps}}$ for all participating schools. 

\begin{figure}[t!!]
	\centering
\caption{Estimated Bounds for Change in School Welfare}
	\includegraphics[width=0.9\textwidth]{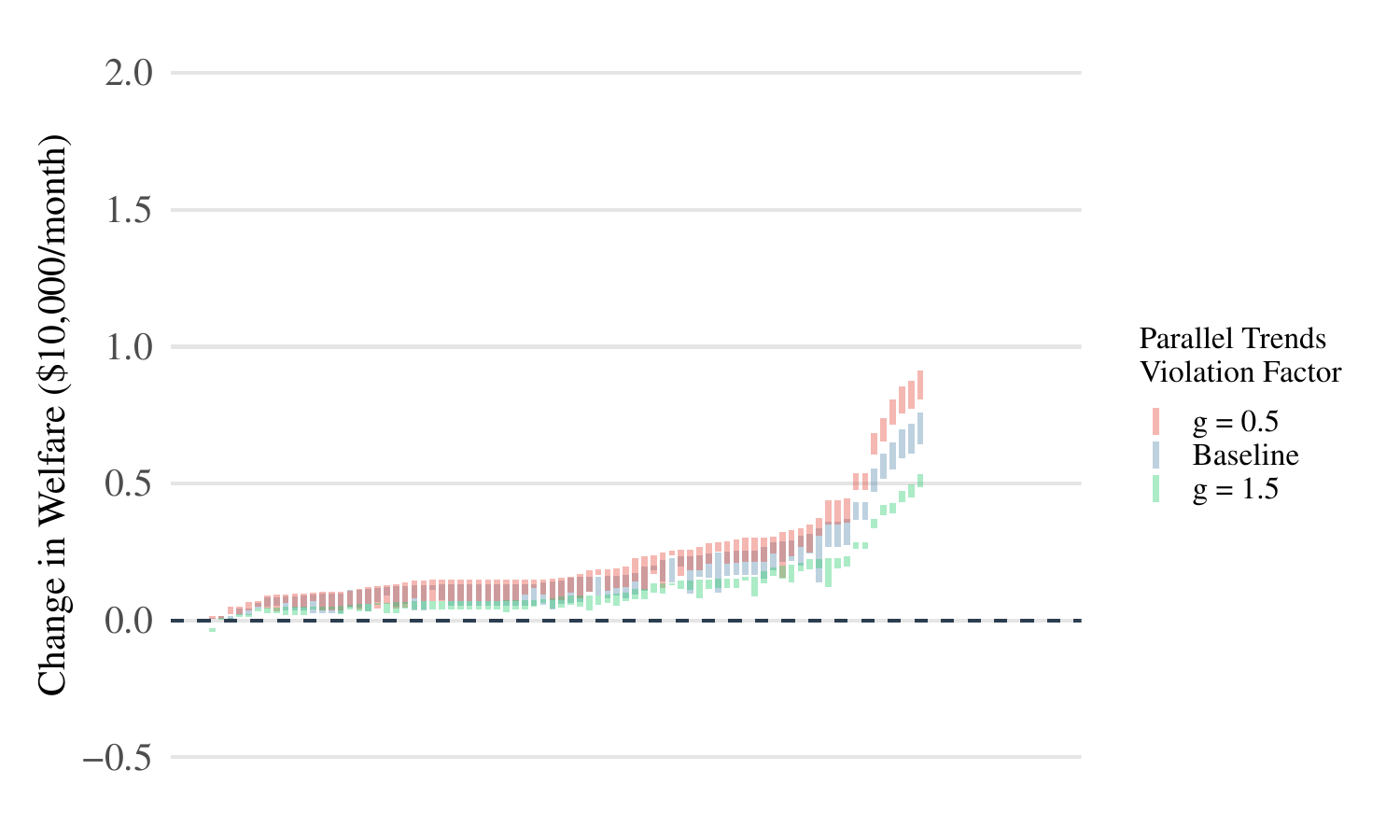}
	\begin{figurenotes}\label{fig:cs_bounds_MANSKI}
		This figure illustrates the bounds on welfare changes (in \$10,000 per month) for participating schools. These bounds are calculated using equation (\ref{eq:cs_bounds}), where we substitute $P_{i1} = P_{i0} + \hat{\beta}^{\text{price}}$ and $Q_{i1} = Q_{i0} + \hat{\beta}^{\text{mbps}}$. The coefficients are derived from equation (\ref{eq:bounds}) using three different values of $g$: 0.5, 1, and 1.5, where $g=1$ denotes the baseline estimate.
		\end{figurenotes}
\end{figure}

Figure \ref{fig:cs_bounds_MANSKI} presents the change in welfare using coefficients defined in (\ref{eq:bounds}) and given in Figure \ref{fig:bounds}. 
Specifically, we consider $g = 0.5$, $g=1$, and $g = 1.5$, representing generous, baseline, and conservative estimates of the program's impact. 
In other words, this figure displays the true effect of the Consortium on participants' welfare for different violations of the parallel trends assumption. 
We find meaningful welfare gains for participating schools, reinforcing the positive impact of the new program.

%% file: conclusion.tex
\section{Conclusion}

In 2014, New Jersey implemented a program for schools to pool demand for broadband internet. 
Our analysis shows that schools can save substantial money by pooling their broadband purchases. 
In particular, we present three main findings: bundling led to lower prices and higher chosen internet speeds; total expenditure savings due to the consortium were between \$1.61 million and \$3.48 million for participating schools, while their existing federal E-rate subsidy was \$2.47 million. There was a substantial increase in school welfare as a result of the consortium.

The success of this program has important policy implications. Helping schools coordinate their buying is at least as effective as the subsidy program in lowering their out-of-pocket expenditure, without requiring taxpayer funding. More broadly, this finding suggests that redesigning how public institutions purchase services may be as effective as traditional subsidies for achieving policy goals.
 
The program addressed two market failures: it helped internet providers mitigate the exposure
problem \citep{milgrom2004putting} and increased competition between providers. We find strong
evidence that reducing bid exposure risk drove most of the savings, with effects concentrated in high fixed-cost Category D services rather than Category A services. Future research should examine how these mechanisms interact. In particular, carefully analyzing competition would require explicitly modeling entry decisions, as the market structure is likely endogenous to procurement design.

Lastly, while bundling may improve immediate procurement outcomes, awarding all the schools in a region to a single ISP could have unintended consequences for market structure. Specifically, ISPs that repeatedly lose bundled auctions may exit the market or reduce their infrastructure investments, potentially leading to reduced competition in the long run. One way to address this could be to use mixed bundling instead of pure bundling, so even smaller ISPs who may be in a better position to serve a few schools (but not all the schools in any one region) would compete, possibly win, and even lower the price further. However, \cite{KimOlivaresWeintraub2014} find that while package bidding allows firms to express cost synergies, ``firms can take advantage of this flexibility by discounting package bids for strategic reasons and not driven by cost synergies,'' generating inefficiencies. This suggests that mixed bundling approaches require careful design, allowing some package combinations while restricting others, creating a tradeoff between promoting competition, managing auction complexity, and balancing short versus long-term benefits. Analyzing this tradeoff is an interesting future research agenda.

%% file: appendix.tex

\section{Appendix: Additional Data Description}
As noted in the main text, there is substantial geographic variation in prices and bandwidth across New Jersey's regions. Table \ref{tab:counts} provides a detailed breakdown of the market, showing the number of contracts signed by each Internet Service Provider (ISP) across the four regions of New Jersey (Northeast, Central, South, and Northwest) in 2014 and 2015. The data reveal significant heterogeneity in the presence of ISPs across regions. Comcast has a strong presence in the Southern region, while Lightpath dominates the Northeast. The Northwest region, being predominantly rural, has notably fewer ISPs and contracts overall, with PenTeleData and NetCarrier being the primary providers.

\input{tables/tab_isp_2014_2015_safe.tex}

Table \ref{tab:sumstats_region} further illustrates this regional variation by comparing key contract characteristics between consortium participants and non-participants across regions. Consistent with the discussion in the main text, the Northwest region exhibits systematically higher prices and lower bandwidth compared to other regions. For instance, in 2014, the average price per Mbps in the Northwest was nearly twice that of other regions, while the average bandwidth was roughly one-fifth of that in other regions. The technology mix also varies substantially across regions, with fiber being less prevalent in the Northwest pre-consortium. However, by 2015, participating schools across all regions showed a marked shift toward fiber connectivity, accompanied by significant price reductions and bandwidth improvements.

\input{tables/tab_sumstats_region_safe.tex}

%% file: tables/tab_isp_2014_2015_safe.tex
\begin{table}[ht!]
\centering
\caption{\label{tab:counts} Aggregated Vendor Data by Region, 2014 and 2015}
\begin{threeparttable}
\begin{tabular}{lrrrr@{\hspace{2em}}rrrr}
\toprule
 & \multicolumn{4}{c}{Pre-Consortium (2014)} & \multicolumn{4}{c}{Post-Consortium (2015)} \\
ISP & N.E. & Cent. & South & N.W. & N.E. & Cent. & South & N.W. \\
\midrule
Comcast* & 17 & 50 & 121 & 10 & 19 & 59 & 139 & 10 \\
Lightpath* & 77 & 33 & 1 & 1 & 99 & 39 & 4 & 1 \\
XO & 7 & 9 & 19 & 2 & 7 & 16 & 33 & 3 \\
Verizon* & 19 & 22 & 15 & 0 & 23 & 13 & 8 & 0 \\
Cablevision* & 21 & 11 & 3 & 1 & 16 & 13 & 3 & 0 \\
PenTeleData* & 0 & 6 & 0 & 16 & 0 & 2 & 0 & 15 \\
NetCarrier & 4 & 1 & 0 & 7 & 4 & 3 & 1 & 8 \\
Windstream & 4 & 0 & 3 & 0 & 6 & 0 & 2 & 0 \\
DNS* & 3 & 6 & 1 & 0 & 5 & 5 & 2 & 0 \\
CenturyLink & 0 & 6 & 1 & 4 & 0 & 4 & 0 & 4 \\
Line Systems & 0 & 1 & 8 & 0 & 0 & 0 & 8 & 0 \\
FiberTech* & 0 & 4 & 5 & 0 & 0 & 2 & 7 & 0 \\
Other & 24 & 10 & 7 & 3 & 16 & 6 & 5 & 6 \\
\midrule
Total & 176 & 159 & 184 & 44 & 195 & 162 & 212 & 47 \\
\bottomrule
\end{tabular}
\begin{tablenotes}[flushleft]
\item \footnotesize \emph{Note:} This table shows the number of school contracts by Internet Service Providers (ISPs)
across four regions of New Jersey in 2014 and 2015. Regions are Northeast (N.E.), Central
(Cent.), South, and Northwest (N.W.). ISPs with a (*) participated in the Consortium. Other
Consortium participants include Aﬀiniti, Cogent, Education Network of America, Lightower,
NexGen, Sunesys, xTel Communications.
\end{tablenotes}
\end{threeparttable}
\end{table}

%% file: tables/tab_sumstats_region_safe.tex
\begin{table}[!htbp]

\caption{\label{tab:sumstats_region} Summary Statistics by Region}
\hspace{-0.3in}\begin{threeparttable}
\begin{tabular}{lllcccccccc}
\toprule
 & & \multicolumn{4}{c}{Pre Consortium (2014)} & \multicolumn{4}{c}{Post Consortium (2015)}\\
Outcome & Participant & Cent. & N.E. & N.W. & South & Cent. & N.E. & N.W. & South \\
\midrule
Participant & Price & 18.64 & 20.59 & 6.71 & 11.57 & 6.35 & 5.50 & 10.92 & 6.83 \\
Non-participant & Price & 12.83 & 18.30 & 25.09 & 16.40 & 11.06 & 14.04 & 16.75 & 12.54 \\
Participant & Bandwidth & 391.97 & 384.00 & 162.50 & 182.15 & 820.68 & 996.15 & 341.67 & 398.00 \\
Non-participant & Bandwidth & 484.11 & 238.87 & 72.77 & 258.50 & 589.26 & 364.43 & 152.15 & 362.30 \\
Participant & Fiber & 0.88 & 0.93 & 0.25 & 0.54 & 0.97 & 1.00 & 1.00 & 0.94 \\
Non-participant & Fiber & 0.71 & 0.74 & 0.50 & 0.62 & 0.75 & 0.82 & 0.64 & 0.69 \\
Participant & Coaxial & 0.09 & 0.07 & 0.75 & 0.46 & 0.03 & 0.00 & 0.00 & 0.06 \\
Non-participant & Coaxial & 0.27 & 0.23 & 0.44 & 0.37 & 0.25 & 0.16 & 0.36 & 0.30 \\
Participant & Other & 0.03 & 0.00 & 0.00 & 0.00 & 0.00 & 0.00 & 0.00 & 0.00 \\
Non-participant & Other & 0.02 & 0.03 & 0.06 & 0.02 & 0.00 & 0.01 & 0.00 & 0.01 \\
Participant & Category D & 0.88 & 0.93 & 0.25 & 0.51 & 0.89 & 0.94 & 0.50 & 0.58 \\
Non-participant & Category D & 0.61 & 0.66 & 0.50 & 0.60 & 0.65 & 0.65 & 0.56 & 0.62 \\
\bottomrule
\end{tabular}
\begin{tablenotes}[flushleft]
\item \footnotesize \emph{Note:} This table presents means of key contract characteristics by region and consortium participation status. Regions are Northeast (N.E.), Central (Cent.), South (South), and Northwest (N.W.).  Price is measured in dollars per Mbps per month, and bandwidth is measured in Mbps. Fiber, Coaxial, and Other are indicator variables for connection type that sum to one within each region-participant-year cell. Category D is an indicator for the service type.
\end{tablenotes}
\end{threeparttable}
\end{table}

%% file: doe.bib
@article{CaplinNalebuff1991,
	author = {Andrew Caplin and Barry Nalebuff},
	journal = {Econometrica},
	number = {1},
	pages = {25-59},
	title = {Aggregation and Imperfect Competition: On the Existence of Equilibrium},
	volume = {59},
	year = {1991}}

@article{rr_honest2023,
	author = {Rambachan, Ashesh and Roth, Jonathan},
	journal = {Review of Economic Studies},
	pages = {rdad018},
	publisher = {Oxford University Press US},
	title = {A More Credible Approach to Parallel Trends},
	year = {2023}}

@article{PyciaTroyan2023,
	author = {Marek Pycia and Peter Troyan},
	journal = {Econometrica},
	number = {4},
	pages = {1495-1526},
	title = {A Theory of Simplicity in Games and Mechansim Design},
	volume = {91},
	year = {2023}}

@article{Li2017,
	author = {Shengwu Li},
	journal = {American Economic Review},
	number = {11},
	pages = {3257-3287},
	title = {Obviously Strategy-Proof Mechanisms},
	volume = {107},
	year = {2017}}

@article{AusubelMilgrom2002,
	author = {Lawrence Ausubel and Paul Milgrom},
	journal = {Advances in Theoretical Economics},
	number = {1},
	pages = {1-42},
	title = {Ascending Auctions with Package Bidding},
	volume = {1},
	year = {2002}}

@article{KatokRoth2004,
	author = {Elena Katok and Alvin E. Roth},
	journal = {Management Science},
	number = {8},
	pages = {1044-1063},
	title = {Auctions of Homogeneous Goods with Increasing Returns: Experimental Comparison of Alternative ``Dutch'' Auctions},
	volume = {50},
	year = {2004}}

@article{RothkopfPekecHarstad1998,
	author = {Michael H. Rothkopf and Aleksandar Peke{\v c} and Ronald M. Harstad},
	journal = {Management Science},
	number = {8},
	pages = {1131-1147},
	title = {Computationally Manageable Combinational Auctions},
	volume = {44},
	year = {1998}}

@article{AveryHendershott2000,
	author = {Christopher Avery and Terrence Hendershott},
	journal = {Review of Economic Studies},
	pages = {483-497},
	title = {Bundling and Optimal Auctions of Multiple Products},
	volume = {67},
	year = {2000}}

@article{SubramaniamVenkatesh2009,
	author = {Ramanathan Subramaniam and R. Venkatesh},
	journal = {Marketing Science},
	number = {2},
	pages = {264-273},
	title = {Optimal Bundling Strategies in Multiobject Auctions of Complements or Substitutes},
	volume = {28},
	year = {2009}}

@article{KrishnaRosenthal1996,
	author = {Vijay Krishna and Robert W. Rosenthal},
	journal = {Games and Economic Behavior},
	pages = {1-31},
	title = {Simultaneous Auctions with Synergies},
	volume = {17},
	year = {1996}}

@article{Chakraborty1999,
	author = {Indranil Chakraborty},
	journal = {Economic Theory},
	number = {3},
	pages = {723-733},
	title = {Bundling Decisions for Selling Multiple Objects},
	volume = {13},
	year = {1999}}

@article{Watt2024,
	author = {Mitchell Watt},
	journal = {Working Paper},
	title = {Concavity and Convexity of Order Statistics in Sample Size},
	url = {https://arxiv.org/pdf/2111.04702},
	year = {2024}}

@article{goolsbee2006impact,
	author = {Goolsbee, Austan and Guryan, Jonathan},
	journal = {The Review of Economics and Statistics},
	number = {2},
	pages = {336--347},
	publisher = {The MIT Press},
	title = {The impact of Internet subsidies in public schools},
	volume = {88},
	year = {2006}}

@article{KagelLienMilgrom2010,
	author = {John H. Kagel and Yuanchuan Lien and Paul Milgrom},
	journal = {American Economic Journal: Microeconomics},
	number = {3},
	pages = {160-185},
	title = {Ascending Prices and Package Bidding: A Theoretical and Experimental Analysis},
	volume = {2},
	year = {2010}}

@techreport{Infrastructure2021,
	author = {{United States Congress}},
	lastchecked = {August 28th, 2024},
	month = {11},
	title = {Infrastructure Investment and Jobs Act},
	url = {https://www.congress.gov/117/plaws/publ58/PLAW-117publ58.pdf},
	year = {2021}}

@techreport{FCC2014,
	author = {{Federal Communications Commission}},
	institution = {{Federal Communications Commission (FCC)}},
	lastchecked = {August 28, 2024},
	title = {Report and Order and Further Notice of Proposed Rulemaking},
	url = {https://docs.fcc.gov/public/attachments/FCC-11-161A1.pdf},
	year = {2014}}

@unpublished{FCC2,
	author = {{Federal Communication Commission}},
	note = {Available from https://www.fcc.gov/general/broadband-deployment-data-fcc-form-477},
	title = {Fixed Broadband Deployment Data from FCC Form 477},
	url = {https://www.fcc.gov/general/broadband-deployment-data-fcc-form-477},
	year = {2014-2015}}

@unpublished{FCC,
	author = {{Federal Communication Commission}},
	note = {Available from https://opendata.usac.org/stories/s/E-Rate-FCC-Form-471-Download-Tool/gifc-3grz/},
	title = {FCC Form 471 E-rate Data},
	url = {https://opendata.usac.org/stories/s/E-Rate-FCC-Form-471-Download-Tool/gifc-3grz/},
	year = {2014-2015}}

@unpublished{NJDRLAP,
	author = {{Educational Services Commission of New Jersey}},
	note = {https://www.escnj.us/co-op-pricing/about-escnj-co-op},
	title = {New Jersey Digital Readiness for Learning and Assessment Program},
	year = {2014-2015}}

@article{GentryKomarovaSchiraldi2023,
	author = {Matthew L. Gentry and Tatiana Komarova and Pasquale Schiraldi},
	journal = {Review of Economic Studies},
	number = {2},
	pages = {852-878},
	title = {Preferences and Performance in Simultaneous First-Price Auctions: A Structural Analysis},
	volume = {90},
	year = {2023}}

@article{roberts2013should,
	author = {Roberts, James W and Sweeting, Andrew},
	journal = {American Economic Review},
	number = {5},
	pages = {1830--1861},
	publisher = {American Economic Association},
	title = {When Should Sellers Use Auctions?},
	volume = {103},
	year = {2013}}

@article{elliott2023market,
	author = {Elliott, Jonathan T and Houngbonon, Georges Vivien and Ivaldi, Marc and Scott, Paul},
	journal = {Journal of Political Economy},
	title = {Market Structure, Investment, and Technical Efficiencies in Mobile Telecommunications},
	year = {Forthcoming}}

@article{DingDugganStarc2025,
	author = {Hui Ding and Mark Duggan and Amanda Starc},
	journal = {Review of Economics and Statistics},
	number = {1},
	pages = {204-220},
	title = {Getting the Price Right? The Impact of Competitive Bidding in the Medicare Program},
	volume = {107},
	year = {2025}}

@article{BajariMcMillanTadelis2009,
	author = {Patrick Bajari and Robert McMillan and Steven Tadelis},
	journal = {Journal of Law, Economics, and Organization},
	number = {2},
	pages = {372-399},
	title = {Auctions Versus Negotiations in Procurement: An Empirical Analysis},
	volume = {25},
	year = {2009}}

@article{CovertSweeney2023,
	author = {Thomas R. Covert and Richard L. Sweeney},
	journal = {American Economic Review},
	number = {628-663},
	title = {Relinquishing Riches: Auctions Versis Informal Negotiations in Texas Oil and Gas Leasing},
	volume = {113},
	year = {2023}}

@article{CBO2020,
	author = {{Congressional Budget Office}},
	journal = {Report},
	lastchecked = {Feb 9, 20204},
	title = {Public-Private Partnerships for Transportation and Water Infrastructure},
	url = {https://www.cbo.gov/publication/56003},
	year = {2020}}

@article{CBO2017,
	author = {{Congressional Budget Office}},
	journal = {Report},
	lastchecked = {Feb 9, 2024},
	title = {A Premium Support System for Medicare: Updated Analysis of Illustrative Options},
	url = {https://www.cbo.gov/system/files/115th-congress-2017-2018/reports/53077-premiumsupport.pdf},
	year = {2017}}

@article{DecarolisPolyakovaRyan2020,
	author = {Francesco Decarolis and Maria Polyakova and Stephen P. Ryan},
	journal = {Journal of Political Economy},
	number = {5},
	pages = {1712-1752},
	title = {Subsidy Design in Privately Provided Social Insurance: Lessons from Medicare Part D},
	volume = {128},
	year = {2020}}

@article{GoereeLindsay2019,
	author = {Jacob K. Goeree and Luke Lindsay},
	journal = {Review of Economic Studies},
	pages = {2230-2255},
	title = {The Exposure Problem and Market Design},
	volume = {87},
	year = {2020}}

@incollection{CantillonPesendorfer2010,
	author = {Estelle Cantillon and Martin Pesendorfer},
	booktitle = {Combinatorial Auctions},
	chapter = {22},
	editor = {Peter Cramton and Yoav Shoham and Richard Steinberg},
	publisher = {MIT Press},
	title = {Auctioning Bus Routes: The London Experience},
	year = {2010}}

@article{FoxBajari2013,
	author = {Jeremy T. Fox and Patrick Bajari},
	journal = {American Economic Journal: Microeconomics},
	number = {1},
	pages = {100-146},
	title = {Measuring the Efficiency of an FCC Spectrum Auction},
	volume = {5},
	year = {2013}}

@article{AusubelCramtonMcAfeeMcMillan1997,
	author = {Lawrence Ausubel and Peter Cramton and R. Preston McAfee and John McMillan},
	journal = {Journal of Economics and Management Strategy},
	pages = {497-527},
	title = {Synergies in Wireless Telephony: Evidence from the Broadband PCS Auctions},
	volume = {6},
	year = {1997}}

@article{KimOlivaresWeintraub2014,
	author = {Sang Won Kim and Marcelo Olivared and Gabriel Y. Weintraub},
	journal = {Management Science},
	number = {5},
	pages = {1180-1201},
	title = {Measuring the Perfermance of Large-Scale Combinatorial Auctions: A Structural Estimation Approach},
	volume = {60},
	year = {2014}}

@article{AgarwalLiSomaini2023,
	author = {Nikhil Agarwal and Pearl Z. Li and Paulo Somaini},
	journal = {NBER Working Paper 31868},
	title = {Identification using Revealed Preferences in Linearly Separable Models},
	year = {2023}}

@incollection{CapliceSheffi2010,
	author = {Chris Caplice and Yossi Sheffi},
	booktitle = {Combinatorial Auctions},
	chapter = {21},
	editor = {Peter Cramton and Yoav Shoham and Richard Steinberg},
	publisher = {MIT Press},
	title = {Combinatorial Auctions for Truckload Transportation},
	year = {2010}}

@article{Beresteanu2005,
	author = {Aire Beresteanu},
	journal = {RAND Journal of Economics},
	number = {4},
	pages = {870-889},
	title = {Nonparametric Analysis of Cost Complementarities in the Telecommunication Industry},
	volume = {36},
	year = {2005}}

@article{xiao2022license,
	author = {Xiao, Mo and Yuan, Zhe},
	journal = {American Economic Journal: Microeconomics},
	number = {4},
	pages = {420--464},
	publisher = {American Economic Association 2014 Broadway, Suite 305, Nashville, TN 37203-2425},
	title = {License Complementarity and Package Bidding: Us Spectrum Auctions},
	volume = {14},
	year = {2022}}

@book{milgrom2004putting,
	author = {Milgrom, Paul Robert},
	publisher = {Cambridge University Press},
	title = {Putting Auction Theory to Work},
	year = {2004}}

@article{kang2022robust,
	author = {Kang, Zi Yang and Vasserman, Shoshana},
	institution = {National Bureau of Economic Research},
	journal = {American Economic Review},
	number = {8},
	pages = {2449-2487},
	title = {Robust Bounds for Welfare Analysis},
	volume = {115},
	year = {2025}}

@article{manski2018right,
	author = {Manski, Charles F and Pepper, John V},
	journal = {Review of Economics and Statistics},
	number = {2},
	pages = {232--244},
	publisher = {MIT Press},
	title = {How Do Right-to-Carry Laws Affect Crime Rates? Coping with Ambiguity Using Bounded-Variation Assumptions},
	volume = {100},
	year = {2018}}
